\documentclass[preprint,aps,floatfix,nofootinbib,tightenlines,showpacs]{revtex4}
\usepackage{dcolumn}%
\usepackage{amssymb}%

\begin{document}


\def\BIGSKIP{\bigskip\noindent}
\def\MEDSKIP{\medskip\noindent}
\def\SMALLSKIP{\smallskip\noindent}
\def\half{\hbox{$1\over2$}}
\def\halfarea{{\cal A}_{1/2}}
\def\quarter{\hbox{$1\over4$}}
\def\ket#1#2{\vert{#1}\enspace{#2}\rangle}
\def\dee{\partial}
\def\prod{{\cal M}}
\def \INFZ{{\rm Inf}[(\gamma/Z)_{\rm L},\nu]}
\def \SSQW{\sin^2\theta_W}
\def \SSQWSQ{\sin^4\theta_W}
\def \up{{\uparrow}}
\def \down{{\downarrow}}
\def \a{{(\uparrow)}}
\def \b{{(\downarrow)}}
\def\bz{\beta_{{}_Z}}
\def \mt{m_t}
\def \mw{m_{{}_W}}
\def \bt{\beta_t}
\def \bw{\beta_{{}_W}}
\def \bb{\beta_b}
\def \gt{\gamma_t}
\def \gb{\gamma_b}
\def \gw{\gamma_{{}_W}}
\def \ZHtrans{$Z\HIGGS$-transverse}
\def \ZHlong{$Z\HIGGS$-longitudinal}
\def \BETA{\bz}
\def \GAMMA{\gamma_{{}_Z}}
\def \GINV{\sqrt{1{-}\BETA^2}}
\def \ns{\enspace}
\def \ts{\thinspace}
\def \nts{\negthinspace}
\def \longitudinal{longitudinal}
\def \amp{{\cal A}}
\def \Ihat{\widehat{\cal I}}
\def \thetas{\theta^{*}}
\def \T{{\cal S}}
\def \G{{\cal G}}
\def \eebar{e^{+}e^{-}}
\def \WWbar{W^{+}W^{-}}
\def \ginv{\gamma_{{}_Z}^{-1}}
\def \GeV{{\rm \enspace GeV}}
\def \beq{\begin{equation}}
\def \eeq{\end{equation}}
\def \beqa{\begin{eqnarray}}
\def \eeqa{\end{eqnarray}}
\def\cphi{c_\varphi}
\def\sph{s_\varphi}
\def\cph{c_\varphi}
\def \cxi{c_\xi}
\def \sxi{s_\xi}
\def \cxip{c_{\xi'}}
\def \sxip{s_{\xi'}}
\def \cxihalf{c_{\xi/2}}
\def \sxihalf{s_{\xi/2}}
\def\xithreehalf{{{3\xi}\over{2}}}
\def\xipthreehalf{{{3\xi'}\over{2}}}
\def\xihalf{{{\xi}\over{2}}}
\def\xiphalf{{{\xi'}\over{2}}}
\def \cchi{c_\chi}
\def \schi{s_\chi}
\def \cth{c_\theta}
\def \sth{s_\theta}
\def\DEE{\mathfrak{D}}
\def\ECKS{{\cal Z}}
\def\WHY{{\cal Y}}
\def\Em{{\cal M}}
\def\quarter{\hbox{$1\over4$}}
\def\half{\hbox{$1\over2$}}
\def\threehalf{\hbox{$3\over2$}}
\def\Vector#1{\overrightarrow{#1}}
\def\braket#1#2{\langle #1 + \vert\thinspace #2 \thinspace\thinspace - \rangle}
\def\tekarb#1#2{\langle #1 - \vert\thinspace #2 \thinspace\thinspace + \rangle}
\def\bra#1#2{\langle #1 \thinspace #2 \vert}
\newfam\cyrilfam
\font\tencyril=wncyr10 scaled \magstep1
\textfont\cyrilfam=\tencyril
\def\cyr{\fam\cyrilfam\tencyril}
\def \eff{{\cyr{f}}}


\preprint{Fermilab--Pub-09-662-T}

\title{Spin Correlation Effects in\\
Top Quark Pair Production at the LHC}
\author{Gregory Mahlon}%
\email{gdm10@psu.edu}
\affiliation{Penn State Mont Alto \\
1 Campus Drive, Mont Alto, PA 17237, USA \\ 
}
\author{Stephen J. Parke}%
\email{parke@fnal.gov}
\affiliation{Theoretical Physics Department\\
Fermi National Accelerator Laboratory \\
P.O. Box 500, Batavia, IL 60510, USA \\
}
\date{January 17, 2010}
\begin{abstract}
\vspace{0.5cm}
At a 14 TeV proton-proton collider, the Large Hadron Collider (LHC),
we show that top quark pair
production is dominated at low invariant mass by 
the fusion of two like-helicity gluons, producing top quark 
pairs in the left-left or right-right helicity 
configurations.  Whereas, at higher 
invariant mass the production is dominated by the fusion of
unlike-helicity gluons, producing
top quark pairs in the up-down or down-up
off-diagonal configurations, identical to top quark pair 
production via quark-antiquark annihilation. 
We study in detail the low invariant mass region, 
and show that the spin correlations can be easily 
observed in this region by looking at the
distribution of the difference in the azimuthal 
angles, $\Delta \phi$, of the dileptons decay products 
of the top quarks in the laboratory frame.   Due to the large cross section for top pair 
production at the LHC, even with a cut requiring that the
invariant mass of the top quark pair be less than 400 GeV, the 
approximate yield would be $10^4$ di-lepton (e, $\mu$) events 
per fb$^{-1}$ before detector efficiencies are applied.  Therefore, 
there is ample statistics to form the $\Delta \phi$ distribution of 
the dilepton events, even with the invariant mass restriction.  
We also discuss possibilities for observing these spin correlations in
the lepton plus jets channel.
\end{abstract}
\maketitle


\vfill\eject\section{Introduction}

Since the discovery of the top quark 
with a mass between 150 and 200 GeV\footnote{The current 
Tevatron-averaged, best-fit value for the top quark mass is
$173.1 \pm 1.3 ~{\rm GeV}$, see Ref.~\cite{canelli}.}
by the CDF and D0 experiments 
at Fermilab in 1995, numerous authors,
see \cite{mahlon and parke}--\cite{Bernreuther:2004jv},
have asked the question
``Can the spin correlations in 
top quark pair production be observed?'' 
This is a valid question 
since the top quark lifetime in the Standard Model is very short 
compared to the spin de-correlation time for such a heavy quark.
In particular, 
$\Gamma_T \sim G_F m_t^3 >> \Lambda^2_{QCD}/m_t$ where 
$\Gamma_T$ is the total width of the top quark, $m_t$ is mass of 
top quark, $G_F$ is the Fermi constant and $\Lambda_{QCD}$ 
is the QCD scale,
thus the top quark decays before QCD interactions have the opportunity
to appreciably affect its spin. The angular distribution of the top 
quark decay products in $t \rightarrow W^+ +b$ followed 
by $W^+ \rightarrow l^+ +\nu$ or $\bar{d} +u$  are correlated 
with the top spin axis as follows:
\begin{eqnarray}
\frac{1}{\Gamma_T} \frac{d \Gamma}{d \cos \chi_i} 
= (1+\alpha_i \cos \chi_i)/2 \quad \alpha_i=\left\{ \begin{array}{l}
+1.0  \quad \hbox{$l^+$ or $\bar{d}$-quark}\\
-0.31  \quad \hbox{$\bar\nu$ or $u$-quark} \\
-0.41 \quad \hbox{$b$-quark}
  \end{array} \right.
\end{eqnarray}
where $\chi_i$ is the angle between the $i$-th decay product and the 
top quark spin axis in the top quark rest frame.
Clearly, the charged lepton or $d$-quark coming from the decay of 
the $W$-boson are the most correlated with the top quark spin axis. 
For the anti-top, the signs of the $\alpha_i$ coefficients are flipped.
Thus, if the spins of the top are correlated in top quark pair 
production and since the decay products of the tops are correlated with 
the spins then the decay products of the two top quarks are 
correlated.  Since there is no net polarization of the top 
quarks, at least to leading order, the correlation between 
the $i$-th decay product of the top and $\bar{\imath}$-th decay 
product of the anti-top can be expressed by
\begin{eqnarray}
\frac{1}{\sigma_T} 
\frac{d^2 \sigma}{d\cos\chi_i d\cos\bar{\chi}_{\bar{i}}} 
= \frac{1}{4} (1+  
C_{t\bar{t}} ~\alpha_i ~\bar{\alpha}_{\bar{i}}~\cos \chi_i 
\cos \bar{\chi}_{\bar{i}} ).
\end{eqnarray}
with
\begin{eqnarray}
C_{t\bar{t}} \equiv 
\frac{
\sigma_{\uparrow \uparrow} 
+ \sigma_{\downarrow \downarrow} 
-\sigma_{\uparrow \downarrow} 
-\sigma_{\downarrow \uparrow} 
} {
\sigma_{\uparrow \uparrow} 
+ \sigma_{\downarrow \downarrow} 
+\sigma_{\uparrow \downarrow} 
+\sigma_{\downarrow \uparrow} 
} .
\end{eqnarray}
$\sigma_{\uparrow/\downarrow~ \uparrow/\downarrow}$ is the 
production cross section for top quark pairs where the
top quark has spin up or down with respect to the top spin axis 
and the anti-top has spin up or down with respect to the antitop 
spin axis.  Clearly, the right choice for the spin axes of the top 
quark pair is important since a poor choice of spin axes can lead 
to a small value for the correlation parameter, $C_{t\bar{t}}$ and 
hence to small correlations between the decay products of the top 
and the antitop.  For some processes, {\it e.g.}\
$q\bar{q} \rightarrow t \bar{t}$, there exist spin axis choices such 
that the correlation parameter, $C_{t\bar{t}}$, is maximal.   
However, even with these optimal choices, to observe the correlations 
one has to measure the angles $\chi_i$ and $\bar{\chi}_{\bar{\imath}}$ 
between the decay products and the spin axes in the rest frame of 
the top and antitop quarks respectively.  To do this one has to 
reconstruct the top and the antitop rest frames; this is very 
challenging at a hadron collider.   
An obvious question is ``Are there variables that carry the 
signature of the spin correlations which  can be measured in the 
laboratory frame?''  For $q\bar{q} \rightarrow t \bar{t}$ no 
such variables have been found that show a significant difference 
between full spin correlations and no spin correlations.

It has been known for some time that the spin correlations at the LHC
are described very well by the helicity basis at low top-antitop 
invariant mass,  whereas at higher invariant mass, 
counter to one's na\"{\i}ve expectation,
the spin 
correlations are degraded in the helicity basis.
In this paper we explain this phenomenon by showing that at low 
invariant mass 
top quark pair production is dominated by like-helicity gluons,
producing top quark pairs in the left-left or right-right helicity 
configuration, independent of the invariant mass.  
Whereas, at higher 
invariant mass top quark pair production is dominated by  
unlike-helicity gluons, producing top quark pairs in the up-down or 
down-up off-diagonal configuration, identical to that of top quark pair 
production via quark-antiquark annihilation. 
At ultra-high invariant masses, the up-down and down-up 
off-diagonal configurations become the familiar left-right or 
right-left helicity configurations; however, at the LHC only a 
small fraction of the total number of produced top pair events are in 
this ultra-high invariant mass region.  
The fact that the contributions from like and unlike helicity
gluons impart different spin correlations to the top quark
pairs makes $gg\rightarrow t\bar{t}$, which dominates
top quark pair production at the LHC, a much richer process
for analysis than the dominant Tevatron mechanism
$q\bar{q}\rightarrow t\bar{t}$.

We study the low invariant mass region in detail.
In this region,
the like-helicity gluons dominate the production.
We show that the spin correlations can be easily 
observed in this region by looking at the
distribution of the difference in the azimuthal 
angles, $\Delta \phi$, of the dilepton decay products 
of the top quarks.
For top quark pairs with an invariant mass $\leq 400$~GeV, the 
spin correlations give a 40\%  enhancement of this distribution 
at small angles ($\Delta \phi \approx 0$) and a 40\% suppression 
of this distribution at large angles ($\Delta \phi \approx \pi$)
compared to if there were no correlations between the production and 
decay of the top quarks.  About 20\% ($\sim 200$~pb) of the 
total next-to-leading order 
top-antitop quark production cross section passes this 
invariant mass cut so even in the di-lepton channel there are large
numbers of events available to measure this distribution: 
about $10^4$ events per fb$^{-1}$ before detector efficiencies are 
applied.

The outline of this paper is as follows:  in Sect.~\ref{sec:qqbar} 
we review 
what is known about the spin correlations for  
$q\bar{q}\rightarrow t\bar{t}$ 
before taking a closer look at the spin correlations for 
$gg \rightarrow t\bar{t}$ in Section~\ref{sec:gg}.
Section~\ref{sec:phenom} addresses the question 
of how the top quark pair events 
at the LHC populate the scattering angle versus invariant mass plane.
In Section~\ref{sec:decayz} we add the 
decays of the top quarks.
In Sect.~\ref{sec:correl} we address the issue of which angular
distributions are sensitive to the presence or absence of
angular correlations among the $t\bar{t}$ decay products.
We show that at 
low invariant mass the difference in the azimuthal angle of the 
charged leptons carries the signature of spin correlations in the 
laboratory frame.  We also discuss possibilities for observing these 
spin correlations in the lepton plus jets channel.
We summarize our conclusions in Sec.~\ref{sec:conclude}.
Lastly, we  
include an Appendix which outlines the highly efficient method 
used in this paper for calculating the spin amplitudes. It can be 
applied to any $2\rightarrow2$ process with arbitrary spins of the 
final state particles.


\vfill\eject
\section{Review of $q \bar{q}  \rightarrow t \bar{t}$}\label{sec:qqbar}
\subsection{Spin Amplitudes in an Arbitrary 
Basis}\label{sec:gen-basis}

\begin{figure*}[hbt]
\includegraphics{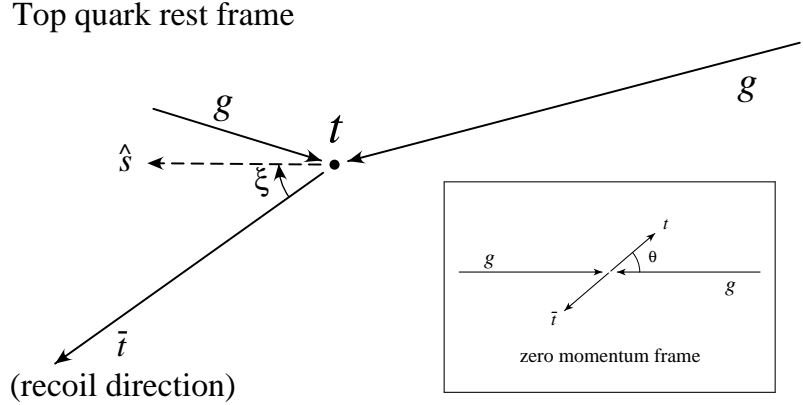}
\vspace*{6cm}
\caption[]{Spin vector for top quark pair production.  The 
direction of the top quark spin vector $s_t$ is given by the angle
$\xi$ in this frame, measured in the clockwise direction
from the recoil direction.  The inset illustrates the situation
in the ZMF, where the top and antitop are produced back-to-back
and the scattering angle $\theta$ from the incoming beam direction.}
\label{fig:s-vector}
\end{figure*}

For a massive particle with momentum, $t$, and spin vector, $s_t$, 
the following relationships are satisfied
\begin{eqnarray}
t^2 =m_t^2, \quad s_t^2=-1, \quad 
{\rm and} \quad t\cdot s_t=0.
\end{eqnarray}
The spin vector $s_t$ is most conveniently defined in the rest frame 
of the massive particle;
in this frame it only can have spatial components.  
Thus, for top-antitop 
quark pair production via quark-antiquark annihilation or gluon-gluon 
fusion, we define the spin vector $s_t$ in the rest frame of the top
quark. Since CP is conserved at tree level for this process, we 
restrict the spin vector to be in the 
scattering plane.  It is convenient to measure the direction of  spin 
vector $s_t$ with respect to the
antitop quark or recoil direction.  Thus we define the unit vector 
$s_t$ such that its direction is at an angle $\xi$,
measured clockwise
with respect to the antitop quark direction 
(see Fig.~\ref{fig:s-vector}).

For the antitop quark we proceed in a similar fashion in 
defining $s_{\bar{t}}$.  Here, instead of using the 
same angle $\xi$ to specify
the direction of the spin vector with respect to the recoil
direction, we use a different 
angle $\xi^\prime$.  This allows for the independent manipulation 
of the antitop and top quark spins at intermediate steps. 
However, in the end, we will set $\xi^\prime =\xi$: 
then, in 
the zero momentum frame (ZMF), the spin 
vectors of the top and antitop quark are back-to-back as expected 
from the symmetry arguments.

It is particularly convenient to use the following combinations 
of $t$ and $s_t$ as follows:
\begin{eqnarray}
t_1 \equiv (t+m_t s_t)/2, 
    \quad {\rm and} \quad 
t_2 \equiv (t-m_t s_t)/2.
\end{eqnarray}
These satisfy
\begin{eqnarray}
t_1^2 = t_2^2 =0, 
\quad  t=t_1+t_2,
\quad {\rm and} \quad 
2t_1\cdot t_2 = m_t^2.
\end{eqnarray}
($\bar{t}_1$ and $\bar{t}_2$ are defined similarly for the antitop 
quark.)
Since $t_1$ and $t_2$ are light-like vectors, the full power of the 
spinor helicity method can
be used in evaluating the amplitudes 
as discussed in Ref.~\cite{Mangano:1990by},
resulting in many simplifications.  
For example, the Dirac spinor for a massive fermion with spin up can be 
written as, see \cite{Kleiss:1985yh}
\begin{eqnarray}
U_\uparrow (t)  = \frac{1+\gamma_5}{2}U(t_1) 
                + e^{i\Psi}\frac{1-\gamma_5}{2}U(t_2).
\end{eqnarray}
This factorization into chiral components, with one depending only 
on $t_1$ and other only $t_2$ (apart from the phase factor 
$e^{i\Psi} \equiv \bar{U}(t_2)\frac{1+\gamma_5}{2} U(t_1)/m_t$), 
is particularly useful.  For our 
purposes in this paper we need only two such spinors:
a $\bar{U}_\uparrow$ spinor for the top quark, and 
a $V_\uparrow$ spinor for the antitop quark.  
These are given in spinor helicity notation by
\begin{eqnarray}
\bar{U}_\uparrow(t) & = 
\langle t_1 + \vert + \displaystyle\frac{\langle t_1+ \vert t_2- \rangle}{m_t} 
\langle t_2 -\vert & = 
 \frac{1}{m_t} ~\langle t_1 + \vert (t+m_t)  \nonumber \\
V_\uparrow(\bar{t}) & = \vert \bar{t}_1-\rangle  
- \vert \bar{t}_2+\rangle 
\displaystyle\frac{\langle \bar{t}_2+ \vert \bar{t}_1- \rangle}{m_t} 
& =(m_t-\bar{t})\vert \bar{t}_1-\rangle~  \frac{1 }{m_t}  
\label{eqn:ubarup+vup}
\end{eqnarray}
The full set of spinors for massive fermions may be found in 
Ref.~\cite{mahlon and parke}.  
In this paper we will show that once you have the amplitudes 
for $gg \rightarrow t_\uparrow \bar{t}_\uparrow$ in terms of the 
spin angles $\xi$ and $\xi^\prime$, you can obtain all of the other
spin combinations for the top and antitop 
by simple algebraic manipulations.


\subsection{Review of the Spin Structure of 
$q \bar{q}  \rightarrow t \bar{t}$}
The spinor structure of the matrix element for  
$q_R \bar{q}_L  \rightarrow t \bar{t}$ can easily be shown 
to be of the form
\begin{eqnarray}
\bar{U}(t) 
\{ \vert q+ \rangle \langle \bar{q}+\vert +\vert \bar{q}- \rangle 
\langle q-\vert \} 
V(\bar{t})
\label{eqn:qqbar-spinor}
\end{eqnarray}
by using the Fierz identities on the current-current structure of the 
matrix element given by the standard Feynman rules.
This can be used to recover the well-known tree level matrix element 
squared for $q_R \bar{q}_L  \rightarrow t \bar{t}$ 
(see Ref.~\cite{Parke:1996pr}):
\begin{eqnarray}
|{\cal A}(q_R \bar{q}_L \rightarrow t_\uparrow \bar{t}_\uparrow \quad 
{\rm and} \quad  t_\downarrow \bar{t}_\downarrow) |^2
& = & |{\cal A}(q_L \bar{q}_R \rightarrow 
t_\uparrow \bar{t}_\uparrow \quad {\rm and} \quad  
t_\downarrow \bar{t}_\downarrow) |^2 \nonumber \\
&\sim &
 (\gamma^{-1}\sin \theta \cos \xi - \cos \theta \sin \xi)^2 \nonumber
\\[0.3cm]
|{\cal A}(q_R \bar{q}_L \rightarrow t_\uparrow \bar{t}_\downarrow 
\quad {\rm or}\quad  t_\downarrow \bar{t}_\uparrow) |^2
&= & |{\cal A}(q_L \bar{q}_R \rightarrow t_\downarrow 
\bar{t}_\uparrow \quad {\rm or}\quad  t_\uparrow \bar{t}_\downarrow) 
|^2 \nonumber \\
& \sim & (\gamma^{-1}\sin \theta \sin \xi + \cos \theta \cos \xi 
\mp 1)^2.
\label{eqn:qq}  
 \end{eqnarray}
Here the same spin angle, $\xi$, has been used for both the top and 
antitop quarks.  From this general result it is clear 
that there is a basis, defined by
\begin{eqnarray}
\tan \xi =\gamma^{-1} \tan \theta,
\end{eqnarray}
which sets the 
$\uparrow \uparrow+\downarrow \downarrow$ component to 
identically zero for all $\beta$, leaving only the  
$\uparrow \downarrow + \downarrow \uparrow$ component.
This basis was first identified by Parke and Shadmi 
in Ref.~\cite{Parke:1996pr} and 
has been called the off-diagonal basis.
It interpolates between the beamline basis 
($\cos \xi =\cos \theta$) at threshold, 
and the helicity basis ($\cos \xi =\pm1$) in the 
ultra-relativistic limit.  In the off-diagonal basis,
Eq.~(\ref{eqn:qq}) becomes
\begin{eqnarray}
&&|{\cal A}(q_R \bar{q}_L \rightarrow t_\uparrow \bar{t}_\uparrow 
\enspace{\rm and}\enspace  t_\downarrow \bar{t}_\downarrow) |^2
= |{\cal A}(q_L \bar{q}_R \rightarrow t_\uparrow \bar{t}_\uparrow 
\enspace{\rm and}\enspace  t_\downarrow \bar{t}_\downarrow) |^2
 = 0 \nonumber  \\[0.3cm]
&&\enspace\thinspace
|{\cal A}(q_R \bar{q}_L \rightarrow t_\uparrow \bar{t}_\downarrow 
\enspace{\rm or}\enspace  t_\downarrow \bar{t}_\uparrow) |^2
= |{\cal A}(q_L \bar{q}_R \rightarrow t_\downarrow \bar{t}_\uparrow 
\enspace{\rm or}\enspace  t_\uparrow \bar{t}_\downarrow) |^2
 \sim  \left(1\mp \sqrt{1-\beta^2 \sin^2 \theta}\right)^2,
\nonumber\\
\label{eqn:qqod}  
\end{eqnarray}
whereas the helicity basis is obtained by setting $\cos \xi = -1$;
in this basis
\begin{eqnarray}
&&|{\cal A}(q_R \bar{q}_L \rightarrow t_R \bar{t}_R 
\enspace{\rm and}\enspace t_L \bar{t}_L) |^2
 = |{\cal A}(q_L \bar{q}_R \rightarrow t_R \bar{t}_R 
\enspace{\rm and}\enspace t_L \bar{t}_L) |^2
\sim \gamma^{-2} \sin^2 \theta \nonumber  \\[0.3cm]
&&\enspace\thinspace |{\cal A}(q_R \bar{q}_L \rightarrow t_R \bar{t}_L 
\enspace{\rm or}\enspace 
t_L \bar{t}_R) |^2
= |{\cal A}(q_L \bar{q}_R \rightarrow t_L \bar{t}_R 
\enspace{\rm or}\enspace  t_R \bar{t}_L) |^2
 \sim  (1\pm \cos \theta )^2.
\end{eqnarray}
Clearly, for $\gamma>>1$, the helicity basis and the Off-Diagonal
basis become identical.
As we will see in the next section, the spin correlations for 
unlike-helicity gluons producing top quark pairs are identical
to those in quark-antiquark annihilation.


\begin{figure*}[hbt]
\includegraphics{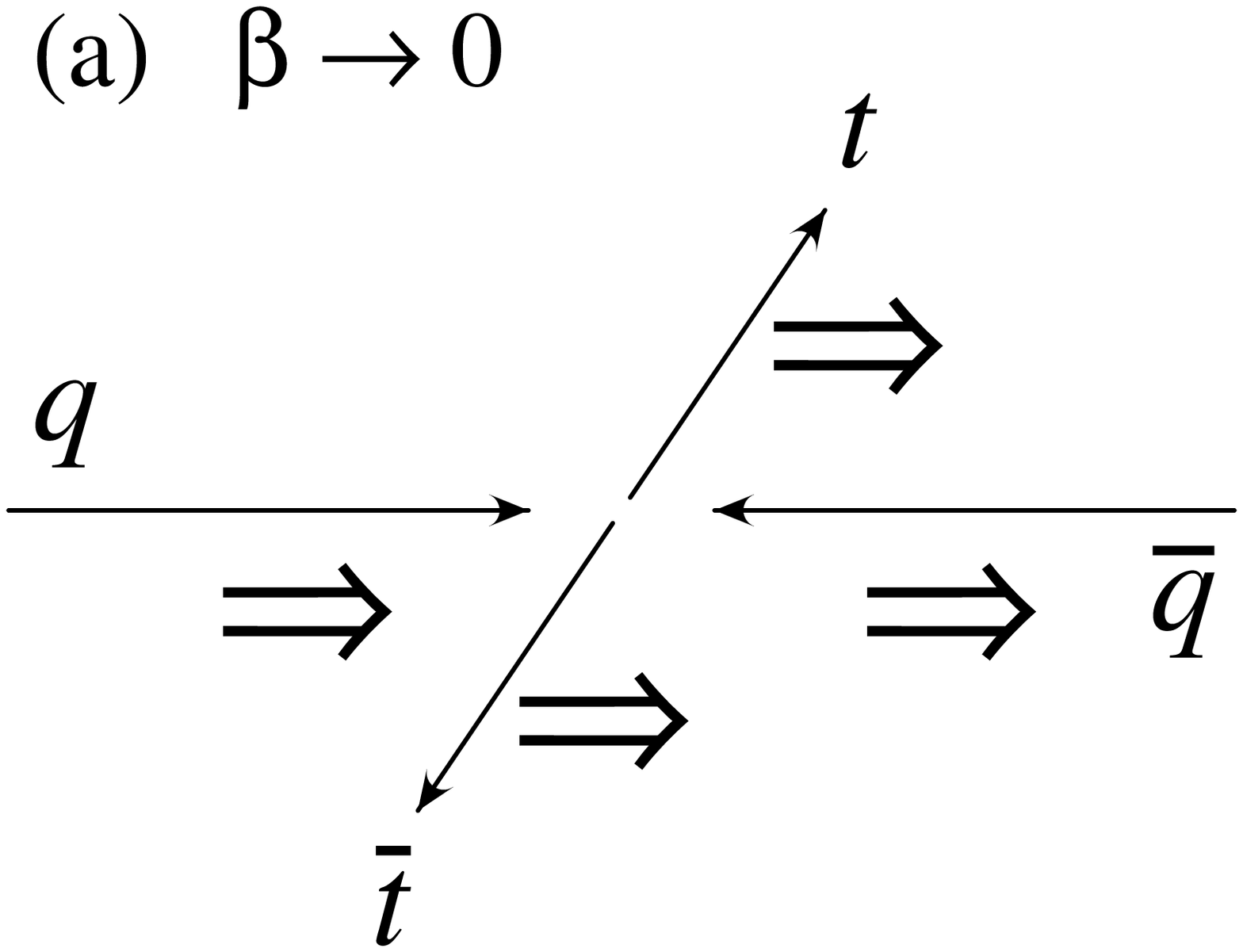}
\includegraphics{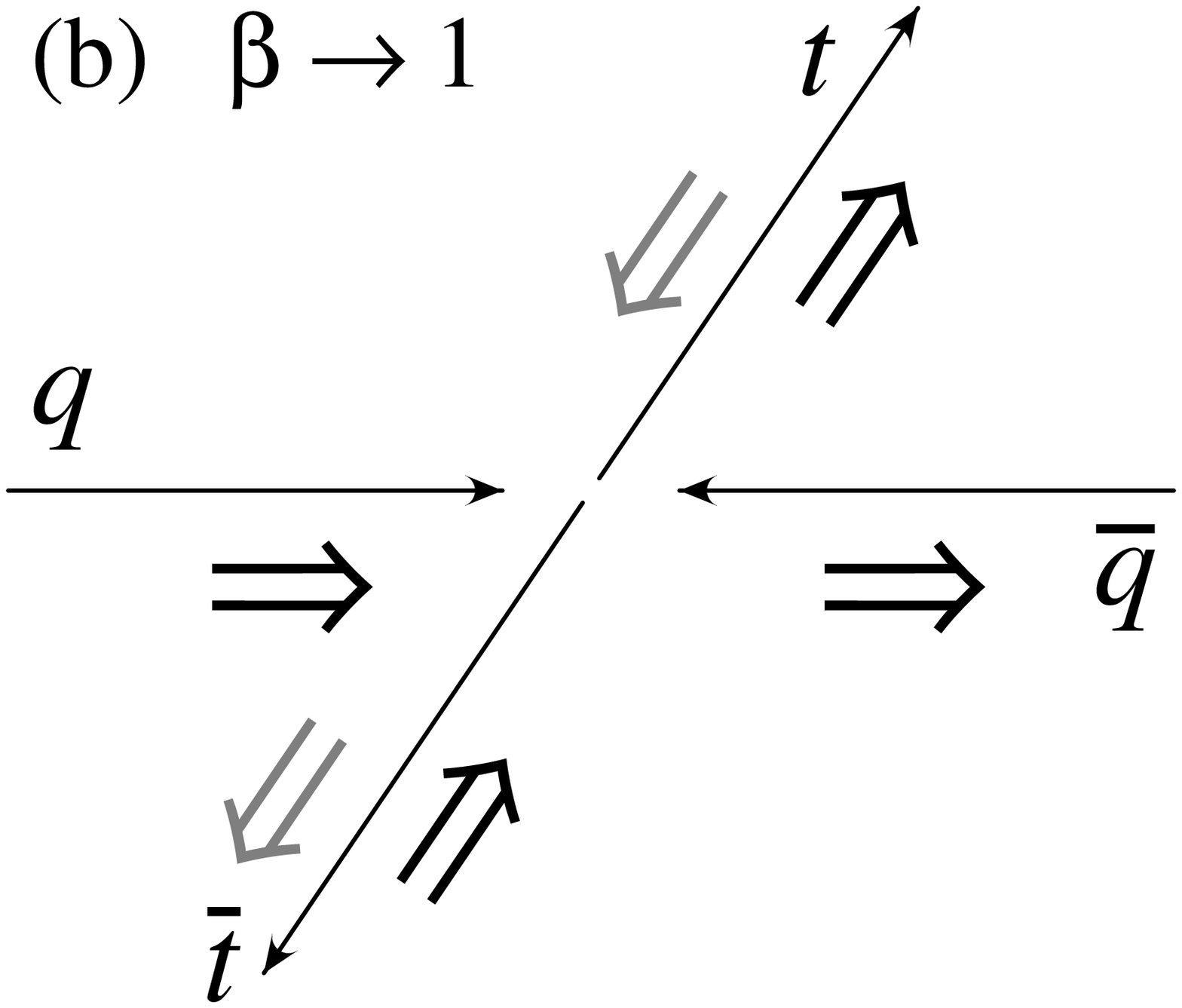}
\vspace*{5.5cm}
\caption[]{The spin configurations for the process 
$q_R \bar{q}_L \rightarrow t \bar{t}$ are best described by the 
off-diagonal basis which interpolates between the beamline basis at 
low $\beta$ to helicity at very high $\beta$ as given by 
Eq.~\protect\ref{eqn:qqod}. 
(a) is the limit $\beta \rightarrow 0$ where the top quark spins are 
aligned in the same direction as the incoming quark spins whereas 
(b) is the limit $\beta \rightarrow 1$ where the helicity state 
$t_R \bar{t}_L$ dominates for scattering angles less than 90 degrees. 
The relative probability of $ t_R \bar{t}_L$ to $t_L \bar{t}_R$ is 
given by $(1+\cos \theta)^2 : (1-\cos \theta)^2$. }
\label{fig:oppo2}
\end{figure*}


\section{A Closer Look at $gg \rightarrow t \bar{t}$}\label{sec:gg}

The tree-level matrix element for $gg \rightarrow t \bar{t}$ can be 
factorized into two terms: one 
depending on the color factors and $t$ and $u$-channel propagators and 
the other depending on the the spin of the
gluons and top quarks, as follows
\begin{eqnarray}
{\cal A}(g_1 g_2 \rightarrow t \bar{t}) = i g^2_s
\left\{ \frac{ [T^{a_1} T^{a_2}]_{\bar{\imath}i}} {(2t\cdot p_1)}
+\frac{[T^{a_2} T^{a_1}]_{\bar{\imath}i}}{(2t\cdot p_2)} \right\} 
M(g_1 g_2 \rightarrow t \bar{t}).
\end{eqnarray}
The reduced matrix element $M(g_1 g_2 \rightarrow t \bar{t})$ is 
symmetric under the interchange of the two gluon momenta
but depends on the the helicity of the gluons and the spin of the top 
and antitop quarks. 

The square of the color-propogator factor, summed over the gluon 
and top quark colors,  is given by 
\begin{eqnarray}
\sum_{color} \left\vert \frac{ [T^{a_1} T^{a_2}]_{\bar{\imath}i}}
{(2t\cdot p_1)} 
+\frac{[T^{a_2} T^{a_1}]_{\bar{\imath}i}}{(2t\cdot p_2)} 
\right\vert^2  = \frac{4}{3}\thinspace
\frac{4(t\cdot p_1)^2 +4(t \cdot p_2)^2 - (t \cdot p_1)(t \cdot p_2)}
{(t\cdot p_1)^2(t\cdot p_2)^2}.
\end{eqnarray}
When evaluated in the ZMF, this sum reduces to the form
\begin{eqnarray}
\sum_{color} \left\vert \frac{ [T^{a_1} T^{a_2}]_{\bar{\imath}i}} 
{(2t\cdot p_1)} 
+\frac{[T^{a_2} T^{a_1}]_{\bar{\imath}i}}{(2t\cdot p_2)} 
\right\vert^2  
= \frac{{\cal Y}(\beta, c_\theta)}{\gamma^4 m_t^4},
\end{eqnarray}
with
\begin{equation}
{\cal Y}(\beta, c_\theta)  = \frac{4}{3} \thinspace
\frac{7+9\beta^2 c^2_\theta}{(1-\beta^2 c^2_\theta)^2}.
\end{equation}
In these expressions
$\beta$ is the ZMF speed of the top quarks and $c_\theta$ is the 
cosine of the ZMF scattering angle $\theta$.  

The reduced matrix 
element for on-mass-shell top quarks, 
$M(g_1 g_2 \rightarrow t \bar{t})$, 
is simply given by
\begin{eqnarray}
M(g_R g_L \rightarrow t \bar{t}) & = &  
\frac{2 ~\langle p_2+ \vert\thinspace t \thinspace\vert 
p_1+\rangle}{2p_1\cdot p_2} \thinspace\bar{U}(t) 
\{\vert p_1+ \rangle \langle p_2 + \vert + \vert p_2- \rangle 
\langle p_1-\vert \} V(\bar{t}) 
\label{eqn:unlike-prod}
\end{eqnarray}
for unlike helicity gluons
and by
\begin{eqnarray}
M(g_R g_R \rightarrow t \bar{t}) & = &  
2m_t \frac{\langle p_1-\vert p_2+\rangle}
{\langle p_1+\vert p_2-\rangle} \thinspace\bar{U}(t) \gamma_L 
V(\bar{t}) \quad {\rm where} \quad \gamma_{L,R} 
\equiv \frac{1}{2}(1\mp \gamma_5) 
\label{eqn:like-prod}
\end{eqnarray}
for like-helicity gluons.
Note the similarity in the spinor structure for 
$g_R g_L \rightarrow t \bar{t}$ and 
$q_R \bar{q}_L  \rightarrow t \bar{t}$. 
Also, the spinor structure for 
$g_R g_R \rightarrow t \bar{t}$ is particularly simple,
$ \bar{U}(t) \gamma_L V(\bar{t})$; it contains no $s$-channel pole. 
In a later section of this paper, we use these two expressions 
to give a simple analytic 
expression for $gg \rightarrow t \bar{t}$ including the decay of the 
two top quarks.  However, in the next section we will evaluate 
these expressions using the 
spinors for polarized top quarks given in Eq.~(\ref{eqn:ubarup+vup}).


\subsection{Unlike-Helicity Gluons}
For unlike-helicity gluons the reduced matrix element 
$M(g_R g_L \rightarrow t_\uparrow \bar{t}_\uparrow)$ is  given by 
\begin{eqnarray}
M(g_R g_L \rightarrow t_\uparrow \bar{t}_\uparrow)
= \frac{2\langle p_2+ \vert t \vert p_1+\rangle}{m_t(2p_1 \cdot p_2)}
\Bigl\{\langle t_1+\vert t_2 \vert p_1+ \rangle 
\langle p_2+ \vert \bar{t}_1- \rangle 
{-}\langle t_1 + \vert p_2-\rangle 
\langle p_1-\vert \bar{t}_2 \vert \bar{t}_1- \rangle \Bigr\},
\nonumber\\
\end{eqnarray}
which, when evaluated in the ZMF using the spin vectors described in 
the previous section, becomes 
\begin{eqnarray}
M(g_R g_L \rightarrow t_\uparrow \bar{t}_\uparrow) & \sim &  
\beta \sin \theta  \{ (1-\cos \theta) \sin (\xi/2) \cos 
(\xi{'}/2) - (1+\cos \theta)\cos( \xi/2) \sin (\xi{'}/2) \nonumber \\ 
&& \enspace\quad \quad + \gamma^{-1} \sin \theta 
[ \cos (\xi/2) \cos (\xi{'}/2) -\sin( \xi/2) \sin (\xi{'}/2)]\}.
\end{eqnarray}

Here the coefficients in front of the products of the $\xi$-dependent
trigonometric functions are the appropriate helicity amplitudes 
whereas the products of the $\xi$-dependent trigonometric functions 
themselves are products of Wigner $d$-functions 
(see Appendix~A). 
The relative signs between the various components of these
expressions are important and care must be taken to make sure they 
are correct.

Using a different spin angle for the $t$ and $\bar{t}$ allows for 
manipulation of the spin of the top independent of the antitop and 
vice versa. Thus, all of the spin amplitudes for
$g_R g_L \rightarrow t \bar{t} $ can be simply obtained from 
$g_R g_L \rightarrow t_\uparrow \bar{t}_\uparrow$ as follows: 
\begin{eqnarray}
&&\vert  M_{\downarrow \uparrow}(\xi, \xi')\vert = 
\vert M_{\uparrow \uparrow}(\xi\pm\pi, \xi') \vert = 
\left|\left(\displaystyle\frac{d}{d \xi/2}\right) 
M_{\uparrow \uparrow}(\xi, \xi') \right|
\label{half:1} \\
&&\vert M_{\uparrow \downarrow}(\xi, \xi') \vert = 
\vert  M_{\uparrow \uparrow}(\xi, \xi'\pm\pi)\vert =  
\left| \left(\displaystyle\frac{d}{d \xi{'}/2} 
\right)M_{\uparrow \uparrow}(\xi, \xi') \right|\\
&&\vert M_{\downarrow \downarrow}(\xi, \xi') \vert = 
\vert M_{\uparrow \uparrow}(\xi\pm\pi, \xi'\pm\pi) \vert= 
\left|\left(\displaystyle\frac{d}{d \xi/2}
\right)  \left(\displaystyle\frac{d}{d \xi'/2} \right)
M_{\uparrow \uparrow}(\xi, \xi')  \right|.
\label{half:3}
\end{eqnarray}
Flipping the spin of a particle is accomplished by 
one of two equivalent methods:
\begin{itemize}
\item{} Addition or subtraction of $\pi$ from the spin angle $\xi$.
\item{} Differentiation of the amplitude with respect to $\xi/2$.
\end{itemize}
A detailed discussion with examples of how to use these techniques for 
arbitrary spins is given in Appendix~A.

At this stage we can make
the spin axes of the top quark pair back-to-back in 
the ZMF by setting $\xi^{'} = \xi$. 
Thus, for unlike-helicity gluons we obtain
\begin{eqnarray}
|{\cal A}(g_R g_L \rightarrow t_\uparrow \bar{t}_\uparrow \enspace 
{\rm and} \enspace  t_\downarrow \bar{t}_\downarrow) |^2
&= &
{\cal Y}(\beta,\theta) ~\beta^2 \sin^2 \theta 
(\gamma^{-1}\sin \theta \cos \xi - \cos \theta \sin \xi)^2,  \\[0.3cm]
|{\cal A}(g_R g_L \rightarrow t_\uparrow \bar{t}_\downarrow \enspace 
{\rm or}\enspace  t_\downarrow \bar{t}_\uparrow) |^2
&= &
{\cal Y}(\beta,\theta) ~\beta^2 \sin^2 \theta(\gamma^{-1}\sin \theta 
\sin \xi + \cos \theta \cos \xi \mp 1)^2,
\label{eqn:unlike}  
 \end{eqnarray}
 and
 \begin{eqnarray}
 |{\cal A}(g_L g_R \rightarrow t_\uparrow \bar{t}_\uparrow \enspace 
{\rm and} \enspace  t_\downarrow \bar{t}_\downarrow) |^2 & = & 
|{\cal A}(g_R g_L \rightarrow t_\uparrow \bar{t}_\uparrow \enspace 
{\rm and} \enspace  t_\downarrow \bar{t}_\downarrow) |^2, \\
  |{\cal A}(g_L g_R\rightarrow t_\downarrow \bar{t}_\uparrow  
\enspace {\rm or}\enspace  t_\uparrow \bar{t}_\downarrow) |^2 & = & 
|{\cal A}(g_R g_L \rightarrow   t_\uparrow \bar{t}_\downarrow \enspace 
{\rm or}\enspace 
  t_\downarrow \bar{t}_\uparrow) |^2,
\end{eqnarray}
with
\begin{eqnarray}
\sum_{\rm all} |{\cal A}(g_R g_L \rightarrow t \bar{t} ) |^2
=  \sum_{\rm all} |{\cal A}(g_L g_R \rightarrow t \bar{t} ) |^2
&= & 2 ~{\cal Y}(\beta,\theta) ~\beta^2 \sin^2 \theta 
(2- \beta^2 \sin^2\theta ).
\end{eqnarray}

\begin{figure*}[hbt]
\includegraphics{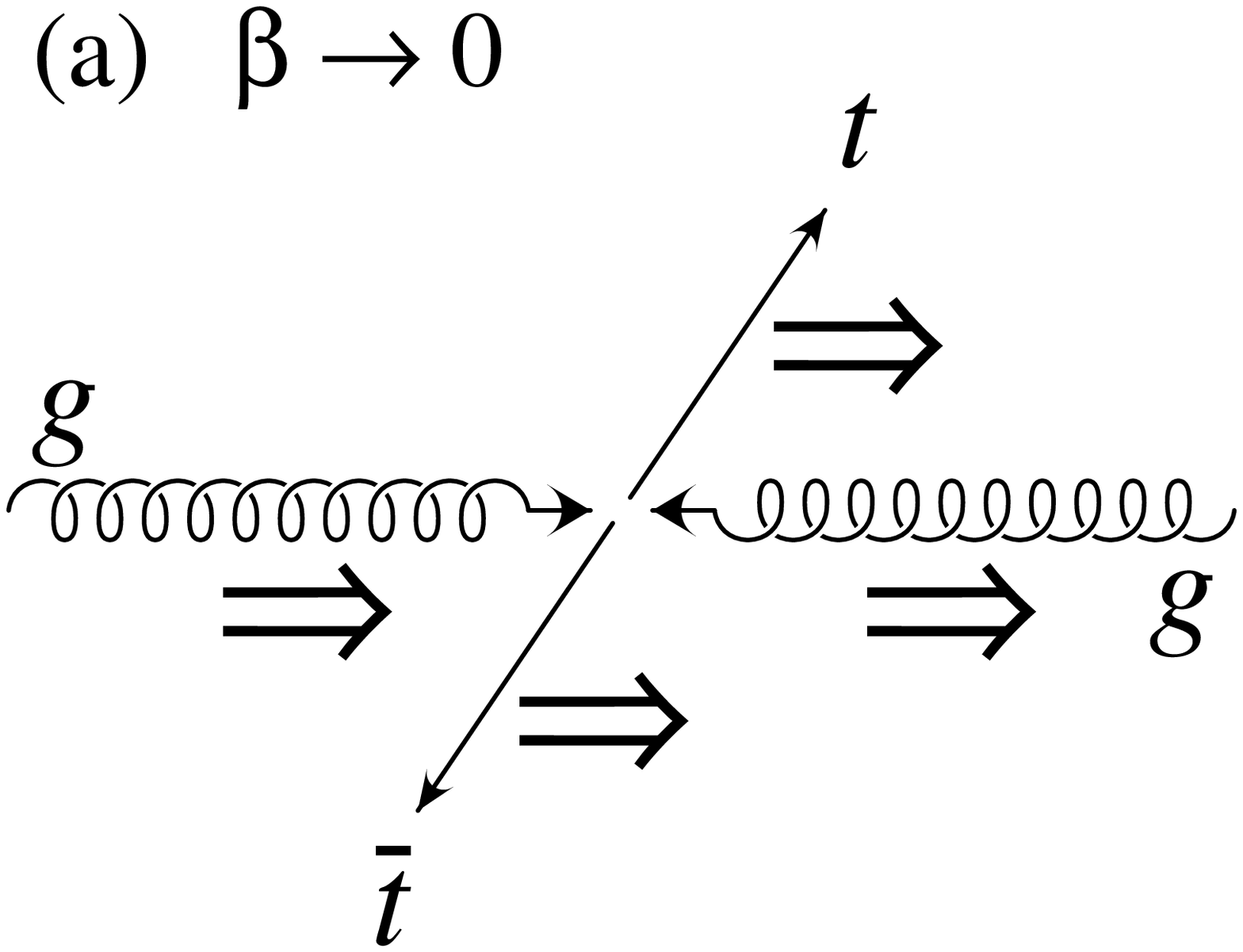}
\includegraphics{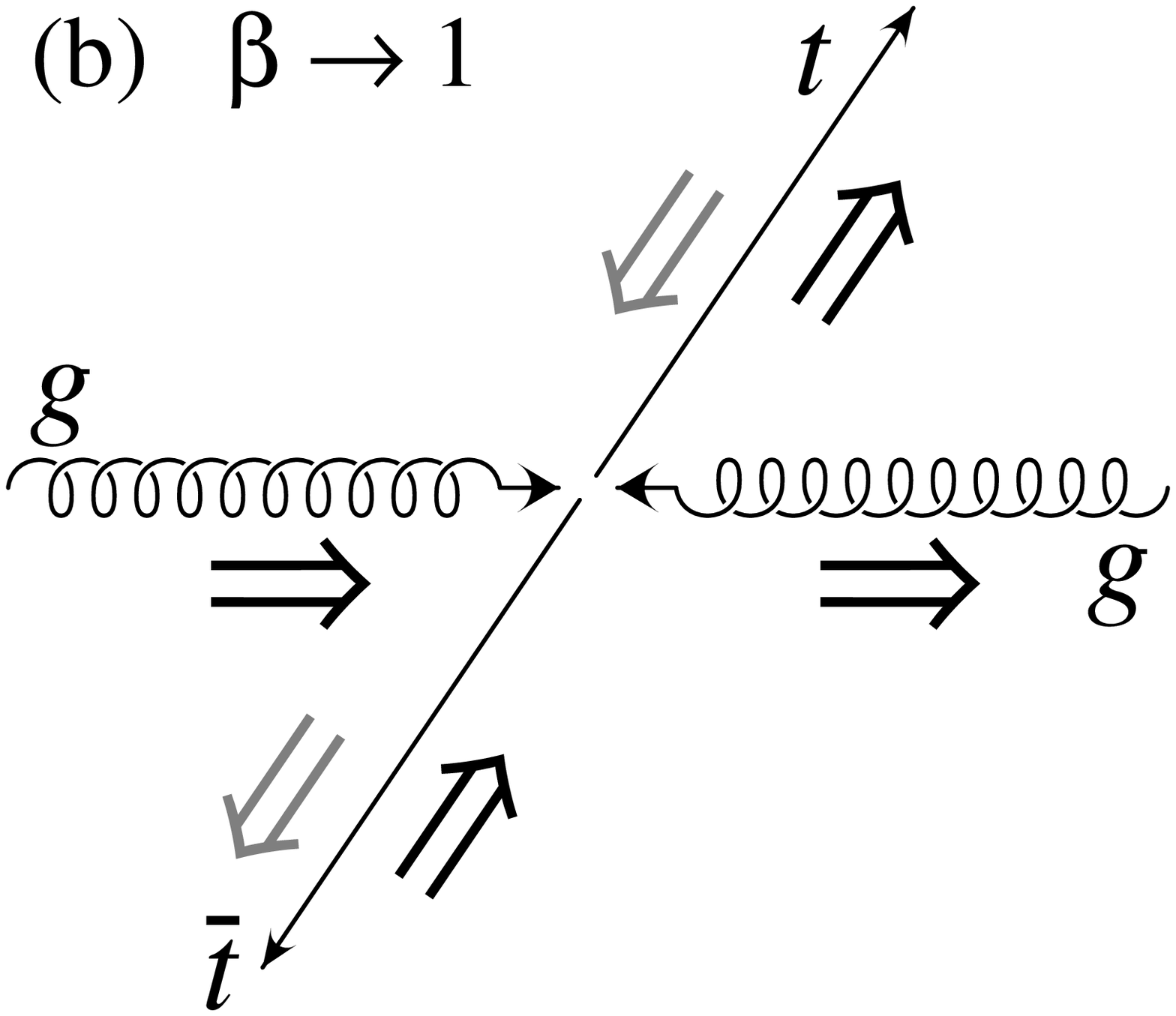}
\vspace{5.5cm}
\caption{The spin configurations for the process 
$g_R g_L \rightarrow t \bar{t}$ are best described by the 
off-diagonal basis, which interpolates between the beamline basis at 
low $\beta$ and the helicity basis at very high $\beta$ (see 
Eq.~\protect\ref{eqn:oppo-od}). As far as the spins of the top 
quarks are concerned, this process, $g_R g_L \rightarrow t \bar{t}$,
is identical to top quark production via quark-antiquark collisions, 
$q_R \bar{q}_L  \rightarrow t \bar{t}$, see 
Fig.~\protect\ref{fig:oppo2}.
(a) illustrates the limit $\beta \rightarrow 0$ where 
the top quark spins are aligned in the same direction as the 
incoming gluon spins whereas 
(b) illustrates the limit $\beta \rightarrow 1$ where the 
helicity state 
$t_R \bar{t}_L$ dominates for scattering angles less than 90 degrees. 
The relative probability of $ t_R \bar{t}_L$ to $t_L \bar{t}_R$ is 
given by $(1+\cos \theta)^2 : (1-\cos \theta)^2$. }
\label{fig:oppo3}
\end{figure*}

As in $q \bar{q} \rightarrow t \bar{t}$, a great simplification 
occurs for the off-diagonal basis~\cite{Parke:1996pr},  
$\tan \xi = \gamma^{-1} \tan \theta$,
where
\begin{eqnarray}
 |{\cal A}(g_L g_R \rightarrow t_\uparrow \bar{t}_\uparrow \enspace 
{\rm and} \enspace  t_\downarrow \bar{t}_\downarrow) |^2  =  
|{\cal A}(g_R g_L \rightarrow t_\uparrow \bar{t}_\uparrow \enspace 
{\rm and} \enspace  t_\downarrow \bar{t}_\downarrow) |^2 = 0.
 \end{eqnarray}
The off-diagonal basis is the basis that interpolates from the 
beamline basis at threshold to the helicity bases at 
ultra-relativistic energies for the  $q \bar{q} \rightarrow t \bar{t}$ 
process.
Thus, for unlike-helicity gluons we have a very similar situation to 
that of $q \bar{q} \rightarrow t \bar{t}$:
the only non-zero amplitudes are given by
\begin{eqnarray}
  |{\cal A}(g_R g_L \rightarrow   
t_\uparrow \bar{t}_\downarrow \enspace 
{\rm or}\enspace 
  t_\downarrow \bar{t}_\uparrow  |^2 & = &   
|{\cal A}(g_L g_R\rightarrow t_\downarrow \bar{t}_\uparrow  \enspace 
{\rm or}\enspace  t_\uparrow \bar{t}_\downarrow |^2 \nonumber  \\
&= &{\cal Y}(\beta,\theta) ~\beta^2 \sin^2 \theta 
\left(1\mp \sqrt{1- \beta^2 \sin^2 \theta}\right)^2,
\label{eqn:oppo-od}
\end{eqnarray}
as illustrated in Fig. \ref{fig:oppo3}.


\subsection{Like-Helicity Gluons}

For like-helicity gluons the reduced matrix element 
$M(g_R g_R \rightarrow t_\uparrow \bar{t}_\uparrow)$ is simply given 
by the following combination of spinor products 
\begin{eqnarray}
M(g_R g_R \rightarrow t_\uparrow \bar{t}_\uparrow)=2m_t 
\frac{ \langle p_1- \vert p_2+ \rangle}{\langle p_1+ \vert p_2- \rangle}
 \langle t_1+ \vert \bar{t}_1 - \rangle,
\end{eqnarray}
which when evaluated in the ZMF using the spin vectors described in 
the previous section is just
\begin{eqnarray}
M(g_R g_R \rightarrow t_\uparrow \bar{t}_\uparrow) & \sim &  
\gamma^{-1} \{ (1-\beta) \cos (\xi/2) \cos (\xi{'}/2) + 
(1+\beta)\sin( \xi/2) \sin (\xi{'}/2) \}.
\end{eqnarray}
Treating these expressions in a manner similar to the unlike-helicity
case discussed in the previous section we obtain
\begin{eqnarray}
&& |{\cal A}(g_R g_R \rightarrow t_\uparrow \bar{t}_\uparrow \enspace 
{\rm or } \enspace  t_\downarrow \bar{t}_\downarrow) |^2
= {\cal Y}(\beta,\theta)~ \gamma^{-2}(1\mp \beta \cos \xi)^2, \\[0.3cm]
&& |{\cal A}(g_R g_R \rightarrow t_\uparrow \bar{t}_\downarrow \enspace 
{\rm and}\enspace  t_\downarrow \bar{t}_\uparrow) |^2
= {\cal Y}(\beta,\theta)~ \gamma^{-2}\beta^2 \sin^2 \xi.
\label{eqn:like}
\end{eqnarray}
Similarly, it is easy to show that for left-handed like-helicity gluons
\begin{eqnarray}
|{\cal A}(g_L g_L \rightarrow  t_\downarrow \bar{t}_\downarrow 
\enspace {\rm or } \enspace   t_\uparrow \bar{t}_\uparrow ) |^2  &= &
|{\cal A}(g_R g_R \rightarrow t_\uparrow \bar{t}_\uparrow 
\enspace {\rm or } \enspace  t_\downarrow \bar{t}_\downarrow) |^2,\\
|{\cal A}(g_L g_L \rightarrow t_\uparrow \bar{t}_\downarrow \enspace 
{\rm and}\enspace  t_\downarrow \bar{t}_\uparrow) |^2
&= &
|{\cal A}(g_R g_R \rightarrow t_\uparrow \bar{t}_\downarrow \enspace 
{\rm and}\enspace  t_\downarrow \bar{t}_\uparrow) |^2.
\end{eqnarray}
Summing over all of the final spins gives
\begin{eqnarray}
\sum_{\rm all} |{\cal A}(g_R g_R \rightarrow t \bar{t} ) |^2 =
\sum_{\rm all} |{\cal A}(g_L g_L \rightarrow t \bar{t} ) |^2
= 2 ~{\cal Y}(\beta,\theta)~ (1-\beta^4),
\end{eqnarray}
independent of the spin axis used for the top quarks.


\begin{figure*}[tbh]
\includegraphics{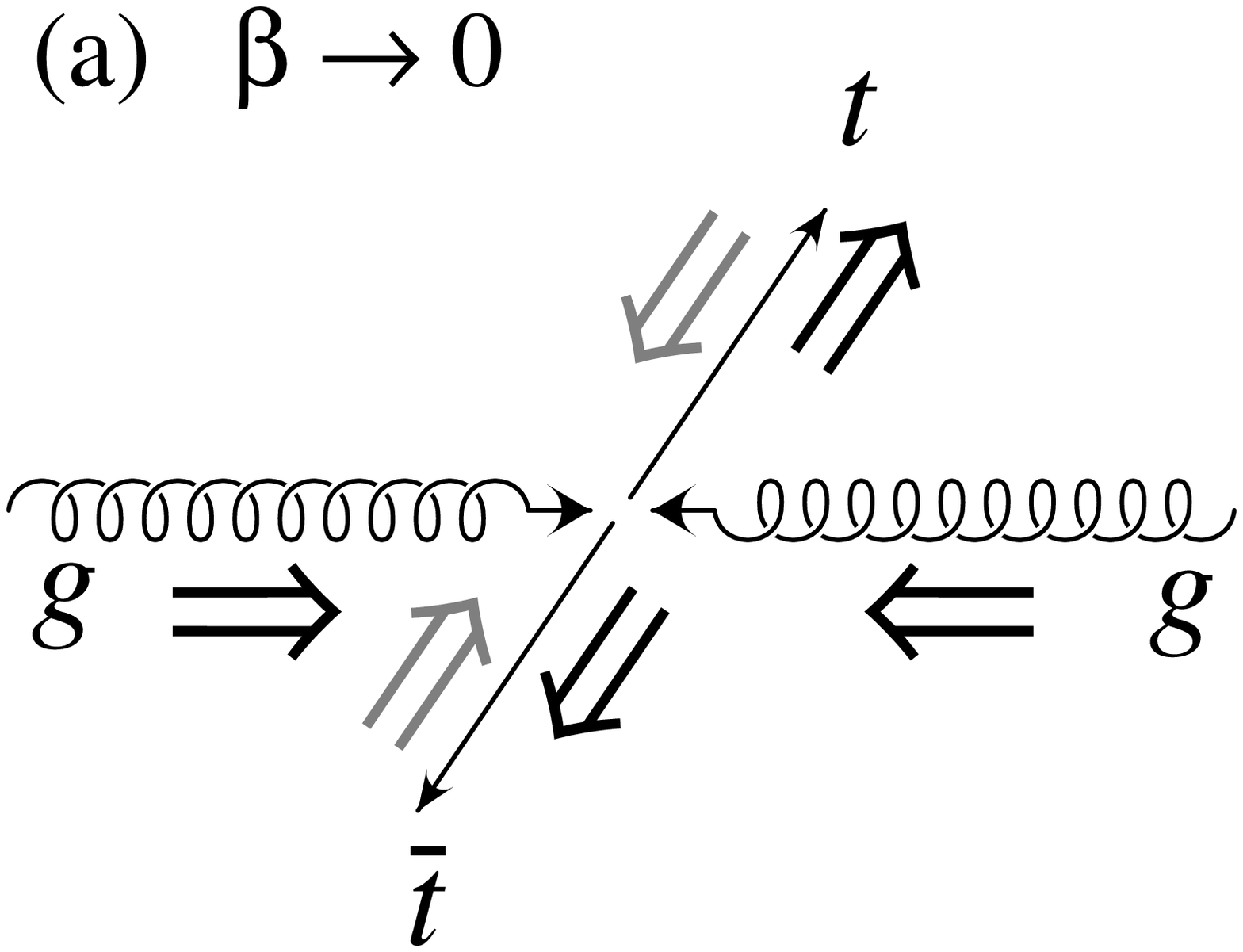}
\includegraphics{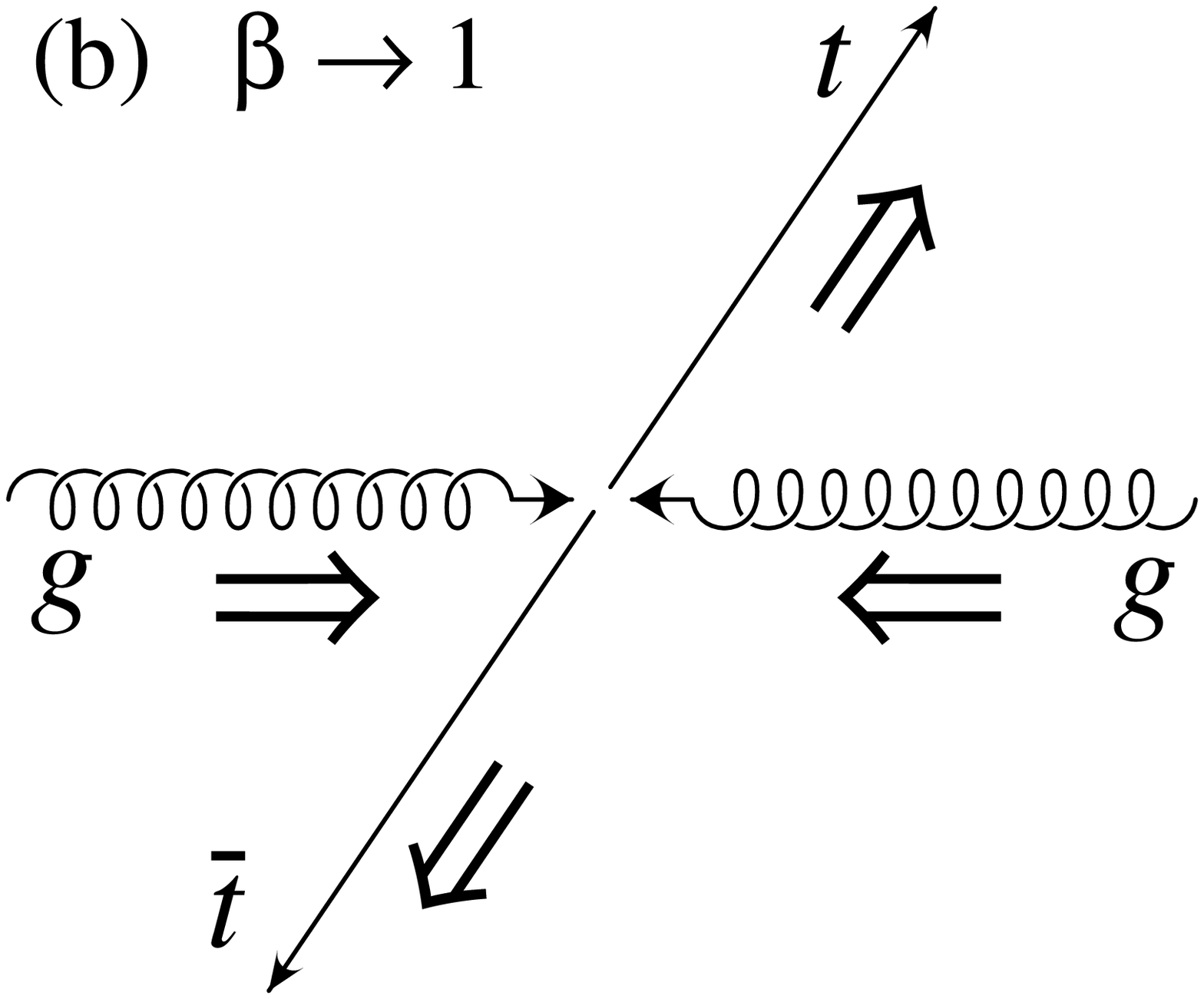}
\vspace*{5.5cm}
\caption[]{The spin configurations for the process 
$g_R g_R \rightarrow t \bar{t}$ are best described by the helicity 
basis for all $\beta$ (see Eq.~\protect\ref{eqn:like-hel}). 
(a) illustrates 
the limit $\beta \rightarrow 0$ where the relative probability 
of $t_R \bar{t}_R$ to  $t_L \bar{t}_L$ is 
$(1+\beta)^2:(1-\beta)^2$ whereas 
(b) illustrates the limit $\beta \rightarrow 1$
where $t_R \bar{t}_R$ completely dominates. 
For $g_L g_L \rightarrow t \bar{t}$,
flip the spins on both the gluons and the top quarks.
}
\label{fig:like}
\end{figure*}

Clearly, a great simplification occurs for like-helicity gluons if 
one uses the helicity basis ($\xi$=0 or $\pi$) for the top quarks. 
In the helicity basis 
\begin{eqnarray}
|{\cal A}(g_L g_L \rightarrow t_R\bar{t}_L \enspace 
{\rm and}\enspace  t_L \bar{t}_R) |^2
= |{\cal A}(g_R g_R \rightarrow t_R \bar{t}_L \enspace 
{\rm and}\enspace  t_L \bar{t}_R) |^2  = 0 
\end{eqnarray}
for {\bf all} values of $\beta$. 
Conventional wisdom states that
helicity provides a simple description for most processes only
at ultra-relativistic energies.  
However, as illustrated in Fig.~\ref{fig:like},
$t\bar{t}$ production from like-helicity gluons is
an exception to this expectation:  in this case, the helicity basis
provieds a simple description for {\bf all}\ $\beta$, with the only
non-zero amplitudes given by
\begin{eqnarray}
|{\cal A}(g_R g_R \rightarrow t_R \bar{t}_R \enspace 
{\rm or } \enspace  t_L \bar{t}_L) |^2  =  |
{\cal A}(g_L g_L \rightarrow  t_L \bar{t}_L 
\enspace {\rm or } \enspace   t_R \bar{t}_R |^2 
=  {\cal Y}(\beta,\theta) ~ \gamma^{-2}(1\pm \beta )^2. 
\label{eqn:like-hel}
\end{eqnarray}
Both the like and unlike gluon helicity amplitudes agree with 
those found in the appendix of Ref.~\cite{hiroshima group}.


\subsection{Combining Like- and Unlike-Helicity Gluons}
At the LHC we must combine the like-helicity and unlike-helicity 
gluon cases since there is no way to 
polarize the incoming gluons.  By looking at Eqns.~(\ref{eqn:unlike})
and~(\ref{eqn:like}), it is 
clear that there is no basis which makes  the top quark spins purely 
$\uparrow \uparrow+\downarrow \downarrow$ OR 
$\uparrow \downarrow + \downarrow \uparrow$ at the LHC because the 
constant term
appears in $\uparrow \uparrow+\downarrow \downarrow$ for 
like-helicity gluons and 
$\uparrow \downarrow + \downarrow \uparrow$ for 
unlike-helicity gluons.

However,  there are regions of the $(\cos \theta, \beta )$ plane for 
which one of the like-helicity
gluon amplitude or the unlike-helicity amplitude dominates.
Along the curve given by 
\begin{eqnarray}
\beta \gamma \sin \theta  = 1 \hbox{ or, equivalently, } 
\beta^2= 1/(2-\cos^2 \theta)
\end{eqnarray}
the like-helicity and the unlike-helicity contribute equally to top 
quark pair production.  On this curve
\begin{eqnarray}
\sum_{\rm all} |{\cal A}(g_R g_R \rightarrow t \bar{t} ) |^2 = 
\sum_{\rm all} |{\cal A}(g_L g_L \rightarrow t \bar{t} ) |^2 =   
\sum_{\rm all} |{\cal A}(g_R g_L \rightarrow t \bar{t} ) |^2 = 
\sum_{\rm all} |{\cal A}(g_L g_R \rightarrow t \bar{t} ) |^2. 
\nonumber
\end{eqnarray}
In the region 
$\beta \gamma \sin \theta <1$ the like-helicity gluon amplitudes
dominate 
the cross section, whereas in the region 
$\beta \gamma \sin \theta >1$ the 
unlike-helicity gluon amplitudes dominate the cross section.
Thus, it is clear that one should use the helicity basis when
$\beta \gamma \sin \theta <<1$ and the off-diagonal basis when 
$\beta \gamma \sin \theta >>1$. In the next section we will optimize 
the basis choice to maximize the spin correlations in the 
intermediate region, $\beta \gamma \sin \theta \sim 1$. 


\subsection{Optimizing the Choice of Spin Basis}
\begin{figure*}[hbt]
\includegraphics{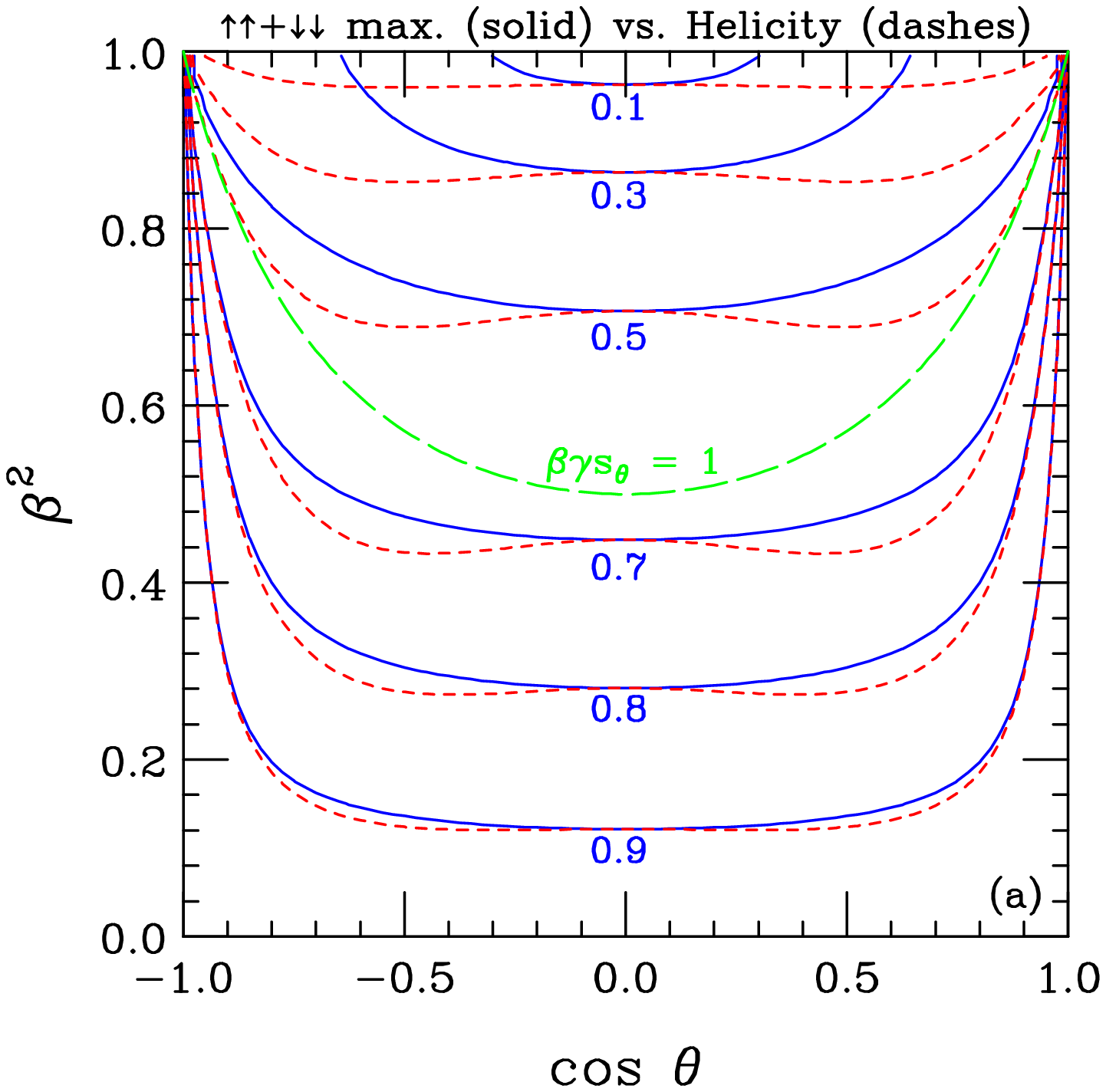}
\includegraphics{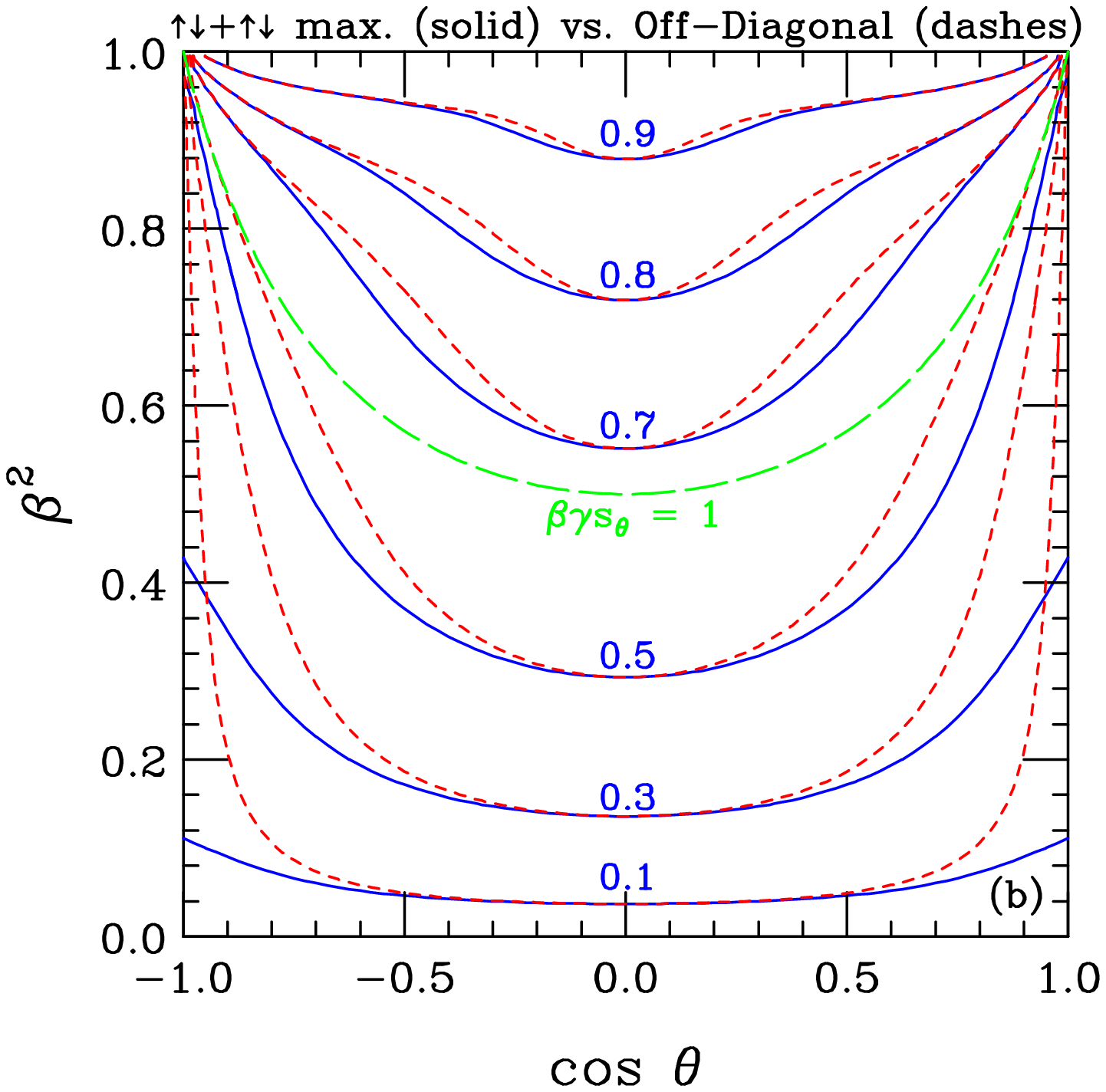}
\vspace{8.5cm}
\caption{Panels (a) and (b) show, in the  $(\cos \theta, \beta^2)$ 
plane, the maximum fractions of  
$\uparrow \uparrow + \downarrow \downarrow$, 
$f_{\rm same}(\theta, \beta)$, and 
$\uparrow \downarrow + \downarrow \uparrow$, 
$f_{\rm oppo}( \theta, \beta)$, respectively (solid contours).  
In both panels the long-dashed line is $\beta \gamma \sin \theta =1$ 
or equivalently $\beta^2=1/(2-\cos^2 \theta)$ below which the 
like-helicity gluons dominate and above the unlike-helicity gluons 
dominate.  In (a) the short-dashed lines are the fractions of LL+RR 
top quark pairs in the helicity basis which below 
$\beta \gamma \sin \theta =1$ are close to the maximum fractions for 
$\uparrow \uparrow + \downarrow \downarrow$, 
$f_{\rm same}(\theta, \beta)$. Whereas in (b) the short-dashed lines 
are the fractions of $\uparrow \downarrow + \downarrow \uparrow$ in 
the Off-Diagonal basis which above 
$\beta \gamma \sin \theta =1$ are close to the maximum fractions of 
$\uparrow \downarrow + \downarrow \uparrow$, 
$f_{\rm oppo}( \theta, \beta)$.
}
\label{fig:fractions}
\end{figure*}

For unpolarized gluons, the fraction of top quark pair events at a 
given point in the $(\cos \theta, \beta)$
plane that have $\uparrow \uparrow$ or $\downarrow \downarrow$ spins 
is
\begin{eqnarray}
f( \theta, \beta) &\equiv& \frac{\sum_{\uparrow \uparrow + 
\downarrow \downarrow} |{\cal A}(g g \rightarrow t \bar{t} ) |^2}
{\sum_{all} |{\cal A}(g g \rightarrow t \bar{t} ) |^2}
\nonumber
\\ &=& 
\frac{  \gamma^{-2}(1+\beta^2 \cos^2 \xi) + \beta^2 \sin^2 \theta 
(\gamma^{-1}\sin \theta \cos \xi - \cos \theta \sin \xi)^2 }
{ ((1-\beta^4) + \beta^2 \sin^2 \theta (2- \beta^2 \sin^2 \theta ) )}. 
\nonumber
\end{eqnarray}
It is a straightforward analytic 
exercise\footnote{The numerical solution was studied 
in Ref.~\cite{Uwer:2004vp}.} to find the extrema of 
this  function with respect to the angle $\xi$. 
The maxima, $f_{\rm same}(\theta, \beta)$, gives the maximum fraction 
of $\uparrow \uparrow + \downarrow \downarrow$ whereas the minima, 
$f_{\rm oppo}( \theta, \beta)$, gives the minimum fraction of
$\uparrow \uparrow + \downarrow \downarrow$ or, equivalently, the 
maximum fraction of
$\uparrow \downarrow + \downarrow \uparrow$. These 
fractions are given by
\begin{eqnarray}
 f_{\{{\rm same,~oppo}\}}(\theta, \beta)  \equiv & & \enspace 
\nonumber \\[0.2cm]
& &  \hspace*{-4cm} \frac{\gamma^{-2}+
\frac{1}{2}\beta^2(\sin^2 \theta \cos^2 \theta + 
\gamma^{-2} \sin^4 \theta + \gamma^{-2}) \left\{
1\pm \sqrt{1- \frac{(2\gamma^{-1} \cos \theta \sin \theta)^2}
{(\sin^2 \theta \cos^2 \theta + 
\gamma^{-2} \sin^4 \theta + \gamma^{-2})^2} } \right\}}
{ ((1-\beta^4) + \beta^2 \sin^2 \theta (2- \beta^2 \sin^2 \theta ) )}.
\end{eqnarray}
Both extrema occur  when $\xi$ satisfies
\begin{eqnarray}
\tan 2 \xi_{\{{\rm same,~oppo}\}} = 
\frac{2 \gamma^{-1} \sin^3 \theta \cos \theta}
{\sin^2 \theta \cos^2 \theta 
-\gamma^{-2} \sin^4 \theta - \gamma^{-2}};
\label{eq:opt}
\end{eqnarray}
they are related as follows: 
$\xi_{\rm oppo} =  \xi_{\rm same} +\pi/2$. 
The contours of  
$f_{\rm same}(\theta, \beta)$ in the 
$(\cos \theta, \beta^2)$ plane are given by the solid lines in 
Fig. \ref{fig:fractions}(a) whereas for 
$f_{\rm oppo}(\theta, \beta)$ see Fig. \ref{fig:fractions}(b).

 
At any given point in the $(\cos \theta, \beta^2)$ 
plane, the basis which 
exhibits the strongest correlations is the one whose spin fraction 
has the 
largest difference from $\frac{1}{2}$.  If 
$|f_{\rm oppo}(\theta, \beta)-1/2|$ is larger than  
$|f_{\rm same}(\theta, \beta)-1/2|$ then one should use 
$\xi_{\rm oppo}$; otherwise, $\xi_{\rm same}$ should be used.  
The condition that must be satisfied for both  
$f_{\rm same}(\theta, \beta)$ and  $f_{\rm oppo}(\theta, \beta)$ to 
have equal difference (but opposite sign) from $1/2$ occurs when
\begin{eqnarray}
 f_{\rm same}(\theta, \beta)+f_{\rm oppo}(\theta, \beta)=1,
\hbox{ or } \beta \gamma \sin \theta =1.
 \end{eqnarray}
Not surprisingly this is the same curve that also separates the 
dominance of the contribution of like-helicity from unlike-helicity 
gluons.  Thus, when $\beta \gamma \sin \theta <1$ the like-helicity 
gluons dominate and $\xi_{\rm same}$ should be used to maximize the  
$\uparrow \uparrow+\downarrow \downarrow$
fraction, whereas if  $\beta \gamma \sin \theta >1$ the 
unlike-helicity gluons dominate and we should use
$\xi_{\rm oppo}$ to maximize the 
$\uparrow \downarrow+\downarrow \uparrow$ fraction; this is 
equivalent to
minimizing the $\uparrow \uparrow+\downarrow \downarrow$ fraction.  
The long-dashed line in both parts of Fig.~\ref{fig:fractions} is the 
curve $\beta \gamma \sin \theta =1$; this  is the demarkation curve 
between maximizing $\uparrow \uparrow + \downarrow \downarrow$ and 
maximizing $\uparrow \downarrow + \downarrow \uparrow$.

It is worthwhile asking the following two questions:
\begin{itemize}
\item{}
In the region dominated by like-helicity gluons 
($\beta \gamma \sin \theta <1$), how much does the
maximum fraction of $\uparrow \uparrow+\downarrow \downarrow$ differ 
from what the helicity basis would give in the same region?  
\item{}
In the region dominated by unlike-helicity gluons
($\beta \gamma \sin \theta >1$), how much does 
the maximum fraction of 
$\uparrow \downarrow+\downarrow \uparrow$ differ 
from what the off-diagonal
basis would give in the same region?  
\end{itemize}
These two questions are also addressed by Fig. \ref{fig:fractions}. 
In (a) we have also plotted (short dashes) the fraction of top 
quark pairs which 
are  LL+RR in the helicity basis and in (b) the fraction that are 
$\uparrow \downarrow + \downarrow \uparrow$ in the off-diagonal 
basis. Clearly, these figures indicate that 
the helicity basis does almost as 
well as the basis which maximizes 
$\uparrow \uparrow+\downarrow \downarrow$ for 
$\beta \gamma \sin \theta <1$ and that the 
off-diagonal basis does almost 
as well as the basis which maximizes 
$\uparrow \downarrow+\downarrow \uparrow$ for 
$\beta \gamma \sin \theta >1$. 

Now that we understand the production dynamics for top quark pair 
production from gluon-gluon fusion we can turn to the question of 
what regions in the $(\cos \theta, \beta^2)$ plane do the top 
quark pair events occur at the LHC.

\section{Phenomenolgy of Top Quark Pair Production at the LHC}
\label{sec:phenom}

 
\begin{figure*}[hbt]
\includegraphics{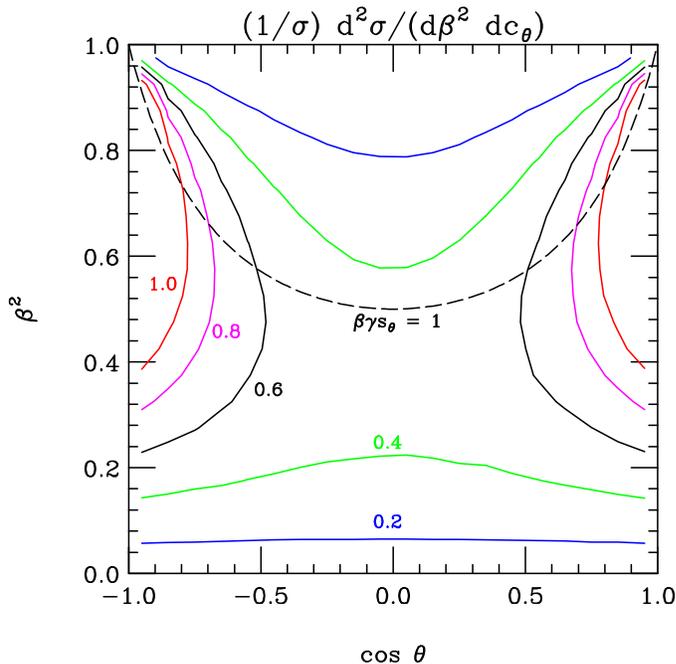}
\vspace{8.5cm}
\caption{The differential cross section 
$(1/\sigma_{\rm tot})~d^2 \sigma /d(\cos\theta)~d\beta^2 $ for top 
quark pair production at the LHC assuming total beam energy of 14 TeV.
The long-dashed line is $\beta \gamma \sin \theta =1$ 
(or $\beta^2=1/(2-\cos^2 \theta)$) is the demarkation line for the 
differential cross section to be dominated by like-helicity gluons 
(below) and unlike-helicity gluons (above). 
}
\label{fig:d2sigma}
\end{figure*}

In Fig. \ref{fig:d2sigma} we plot the differential cross section 
\begin{eqnarray}
\frac{1}{\sigma_{\rm tot}}~\frac{d^2 \sigma }
{d(\cos\theta)d\beta^2 } ~(gg \rightarrow t \bar{t}) \nonumber
\end{eqnarray}
in the $(\cos \theta, \beta^2)$ plane where $\sigma_{\rm tot}$ is 
the total cross section.
This figure gives us the relative distribution of top quark pair 
events in this
plane.  
A breakdown of the fraction of events for like- and unlike-helicity 
gluons 
broken into the appropriate regions in $\beta \gamma s_\theta$ are 
given in Table~\ref{tab:regions}.


\begin{figure*}[tbh]
\includegraphics{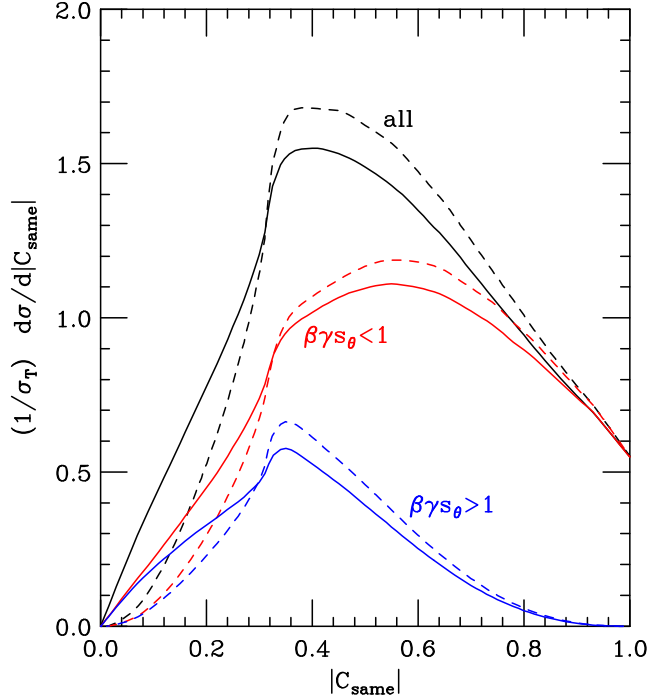}
\vspace*{9.5cm}
\caption{The differential distributions of $\vert C_{\rm same} \vert $, 
$(1/\sigma_{{}_T})~ d\sigma/d \vert C_{\rm same} \vert $.
The dashed curves are for the optimal basis, 
Eqn.~(\protect\ref{eq:opt}), 
whereas the solid curves employ the helicity basis 
when $\beta \gamma s_\theta < 1$
and the off-diagonal basis when $\beta \gamma s_\theta > 1$.
The mean values of $\vert C_{same} \vert $ for the 
different bases and regions are collected in Table~\ref{tab:meanC}.}
\label{fig:Cdistr}
\end{figure*}


%
\begin{table}
\caption{Fraction of $t\bar{t}$ events catagorized by the helicities
of the initial state gluons and location in the $(\cos\theta,\beta^2)$
plane at the LHC with $\protect\sqrt{s}=14$~TeV.}
\begin{center}
\begin{tabular}{lccc}
\hline\hline
 & $\beta\gamma\sth<$1 & $\beta\gamma\sth>$1 & all \\
\hline
$gg$ Like   & 55\% & 10\% & \phantom{1}65\% \\
$gg$ Unlike & 20\% & 15\% & \phantom{1}35\% \\
Total       & 75\% & 25\% & 100\% \\
\hline\hline
\end{tabular}
\end{center}
\label{tab:regions}
\end{table}

Fig. \ref{fig:Cdistr} shows the differential cross section with
respect to $\vert C_{\rm same} \vert$, where 
\begin{eqnarray}
C_{\rm same} \equiv 
2f_{\uparrow \uparrow+\downarrow \downarrow}-1 \hbox{ and } 
C_{\rm oppo} \equiv 
2f_{\uparrow \downarrow+\downarrow \uparrow}-1 
\end{eqnarray} 
are the quantities which control the size of the correlations for any 
given spin basis (note that $C_{\rm oppo}= - C_{\rm same}$).  
In this figure
we have split up the contributions into two pieces: one for 
$\beta \gamma \sin \theta <1$ and the other for
$\beta \gamma \sin \theta \ge 1$ .  For $\beta \gamma \sin \theta <1$,
we show the contribution in the basis which maximizes the 
$\uparrow \uparrow+\downarrow \downarrow$ component as well as the 
helicity basis; whereas for $\beta \gamma \sin \theta \ge 1$, we show 
the contribution in the basis which maximizes 
$\uparrow \downarrow+\downarrow \uparrow$ as well as the off-diagonal 
basis. 
This figure clearly shows that there are only small differences 
between using the best basis and the 
helicity basis
for $\beta \gamma \sin \theta <1$ and the best basis and the 
off-diagonal basis for $\beta \gamma \sin \theta >1$.

At the LHC, 
the total top quark pair production cross section is $\sim$1~nb 
(at next-to-leading order)
giving approximately $10^6~t\bar{t}$ per fb$^{-1}$; therefore,
significant cuts can be made on the data before the statistical 
uncertainties become comparable to the systematic uncertainties.
Thus, we will concentrate on the low $\beta$ region for the rest of 
this paper since in this region like-helicity gluons dominate
the production cross section
and the boost of the top quarks does not mask the spin correlations.


\begin{table}
\caption{Mean values of $\vert C_{\rm same} \vert$  using either the
optimal or appropriate choice of off-diagonal or helicity basis 
for events with different values of $\beta\gamma\sth$ at the
LHC with $\protect\sqrt{s}=14$~TeV.
}
\begin{center}
\begin{tabular}{lccc}
\hline\hline
 Basis & $\beta\gamma\sth<$1 & $\beta\gamma\sth>$1 & all \\
\hline
Optimal basis             & 0.55 & 0.59 & 0.43 \\
Helicity or off-diagonal
& 0.53 & 0.57 & 0.39  \\
\hline\hline
\end{tabular}
\end{center}
\label{tab:meanC}
\end{table}


\newpage
\section{Adding Top Quark Decays}\label{sec:decayz}
 From Eqs.~(\ref{eqn:unlike-prod}) and~(\ref{eqn:like-prod}), 
it is easy to add the decays of the on mass shell top quarks, 
$t \rightarrow b+\bar{e}+\nu$ and
$\bar{t} \rightarrow \bar{b} + \mu +\bar{\nu}$, via the following 
replacements:
\begin{eqnarray}
\bar{U}(t) & \rightarrow & 
{ 
 g_{w}^2
\over
{(2\bar{e}\cdot\nu - m_{w}^2 + i m_{w}\Gamma_{w})
 ( im_t \Gamma_t)}
} {\langle b-\vert \nu+\rangle \langle \bar{e}+\vert(t+m_t)} \nonumber \\
V(\bar{t}) &\rightarrow & 
{(-\bar{t}+m_t) 
\vert \mu +\rangle   \langle  \bar{\nu}+ \vert \bar{b} -\rangle}
{  g_{w}^2 \over
{(2\mu\cdot\bar\nu - m_{w}^2 + i m_{w}\Gamma_{w}) 
 ( im_t \Gamma_t)}
}. 
\label{eqn:add-decays}
\end{eqnarray}
The Fierz identity has been employed in the derivation of these 
replacements.
Thus, the total matrix element squared for top quark production 
and decay via gluon fusion, summed over the colors of the 
incoming gluons and outgoing $b$-quarks, is given by
\begin{eqnarray}
\vert {\cal A}\vert ^2_{RL}  + \vert {\cal A}\vert ^2_{LR}  
= {\cal K}\Biggl\{ 
\frac{ (2 (p_1 \cdot t)(p_2 \cdot t)-m_t^2(p_1 \cdot p_2)}
{(p_1 \cdot p_2)^2} \Biggr\}
\Biggl\{~2(t \cdot \bar{e})(\bar{t} \cdot \mu)
\Bigl[(p_1\cdot t)^2+(p_2 \cdot t)^2 \Bigr] \nonumber \\
-m_t^2\Bigl[~(p_1\cdot p_2) \Bigl( (t \cdot \bar{e})(t\cdot \mu)
+(\bar{t}\cdot \bar{e})(\bar{t} \cdot \mu)-m_t^2(\bar{e} \cdot \mu)\Bigr)
\nonumber \\ 
-2 \Bigl((p_1 \cdot t)(p_1 \cdot \mu) (p_2 \cdot \bar{e}) 
+(p_2 \cdot t)(p_1 \cdot \bar{e})(p_2 \cdot \mu) 
-(p_1 \cdot t)(p_2 \cdot t)(\bar{e} \cdot \mu)\Bigr)~\Bigr]~\Biggr\} \nonumber \\
\label{eqn:unlike-full}
\end{eqnarray}
for unlike-helicity gluons, whereas for like-helicity gluons we have
\begin{eqnarray}
\vert {\cal A}\vert ^2_{RR}  + \vert {\cal A}\vert ^2_{LL}  =  
{\cal K}~m_t^4 \{ (t\cdot \bar{e})(t\cdot \mu)
+(\bar{t}\cdot \bar{e})(\bar{t} \cdot \mu)
-m_t^2 (\bar{e} \cdot \mu)\}.
\label{eqn:like-full}
\end{eqnarray}
The overall factor ${\cal K}$ is given by
\begin{eqnarray}
{\cal K} = {{2^6g_s^4}\over{3}} 
{{g_w^8}\over{(m_t \Gamma_t)^4}}
\left\{ 
\frac{ 4(p_1 \cdot t)^2+4 (p_2 \cdot t)^2 - (p_1 \cdot t)(p_2 \cdot t)}
{(p_1 \cdot t)^2 (p_2 \cdot t)^2} \right\} \qquad\nonumber \\
\qquad\times
{{b\cdot\nu}\over
{ (2\bar{e}\cdot\nu - m_w^2)^2 + (m_w\Gamma_w)^2 }}
{{\bar{b}\cdot\bar\nu}\over
{ (2\mu\cdot\bar\nu - m_w^2)^2 + (m_w\Gamma_w)^2 }}.
\label{eqn:ggKfactor}
\end{eqnarray}
Appendix~B contains the corresponding expressions 
describing  $q\bar{q}\rightarrow t\bar{t}$.

Notice the simplicity of the matrix element squared for like-helicity 
gluons to top quark pairs. Given this simplicity and the fact that 
the like-helicity gluon 
contribution dominated at smaller values of the invariant mass 
of the $t\bar{t}$ system, it is worth exploring whether or not the 
full matrix element enhances the spin correlations in this channel.

The ratio of the correlated to 
uncorrelated\footnote{We call the decay of a top or anti-top quark 
into a $W$-boson and $b$-quark
uncorrelated if this decay is spherical in the top quark rest frame and 
thus independent of the top quark spin.  The $W$-boson is then 
assumed to decay in the usual (fully correlated) manner.
The uncorrelated matrix elements squared are then given by 
$
(\vert {\cal A}\vert^2_{RR} 
+ \vert {\cal A}\vert^2_{LL} )_{uncorr}
 =  {\cal K}~(t\cdot \bar{e})(\bar{t} \cdot \mu) ~\{ m_t^2
(t\cdot \bar{t})\}$
and
$(\vert {\cal A}\vert ^2_{RL}  + \vert {\cal A}\vert ^2_{LR} )_{uncorr}
=  {\cal K} (t\cdot \bar{e})
(\bar{t}\cdot\mu)
\Bigl\{
[2 (p_1 \cdot t)(p_2 \cdot t)-m_t^2(p_1 \cdot p_2)]
 [ (p_1 \cdot t)^2+(p_2 \cdot t)^2 +m_t^2 (p_1 \cdot p_2) ]
/(p_1 \cdot 
p_2)^2\Bigr\} . $}
matrix element squared, ${\cal S}$, for like-helicity gluons 
is given by
\begin{eqnarray}
{\cal S} \equiv 
\frac{(\vert {\cal A}\vert ^2_{RR}  
+ \vert {\cal A}\vert ^2_{LL})_{corr}}
{( \vert {\cal A}\vert ^2_{RR}  
+ \vert {\cal A}\vert ^2_{LL})_{uncorr}}
 & = & \frac{m_t^2 \{ (t\cdot \bar{e})(t\cdot \mu)
+(\bar{t}\cdot \bar{e})(\bar{t} \cdot \mu)-m_t^2 
(\bar{e} \cdot \mu)\} }
{ (t\cdot \bar{e})(\bar{t} \cdot \mu)(t\cdot \bar{t}) } \nonumber \\
& = & \left( \frac{1-\beta^2}{1+\beta^2} \right)  
\left( \frac{(1+\beta^2) + (1-\beta^2)c_{\bar{e}\mu} 
- 2 \beta^2 c_{t\bar{e}} c_{\bar{t}\mu}}
{(1-\beta c_{t\bar{e}})(1-\beta c_{\bar{t}\mu})} \right),
\label{eqn:ratio}
\end{eqnarray}
where the last line is given in the ZMF in terms of speed of the 
tops, $\beta$, and the cosine of the angles between
$t$ and $\bar{e}$ ($c_{t\bar{e}}$), $\bar{t}$ and $\mu$ 
($c_{\bar{t}\mu}$) and $\bar{e}$ and $\mu$ ($c_{\bar{e}\mu}$).
The range of ${\cal S}$ is between (2,0).  At threshold, 
$\beta \rightarrow 0$, the maximum of ${\cal S}$ occurs when the 
charged leptons are parallel, $c_{\bar{e}\mu}=+1$, whereas the 
minimum occurs when the charged leptons are back-to-back, 
$c_{\bar{e}\mu}=-1$, independent of their correlation with the 
top-antitop axis.

For non-zero $\beta$, the maximum (minimum) still occurs when the 
charged leptons are parallel (back-to-back), but they are now 
correlated with the top-antitop axis. 
The fact that the charged leptons are more likely to have their
momenta being parallel rather than back-to-back is what is 
expected for top quark pairs that have spins which are 
anti-aligned, {\it i.e.}\  LL or RR. However, here the enhancement is 
even stronger than what one would na{\"\i}vely expect because the 
interference between LL and RR strengthens the correlation between the 
momenta of the two charged leptons.  This argument suggests
looking at the 
$\Delta R$, $\Delta \eta$ and 
$\Delta \phi$ distributions of the
two charged leptons with a cut on the 
invariant mass of the top-antitop system.


\section{Correlation-Sensitive Angular Distributions}\label{sec:correl}


\subsection{$\Delta \phi$ Distribution in Dilepton Events}
\begin{figure*}[tbh]
\includegraphics{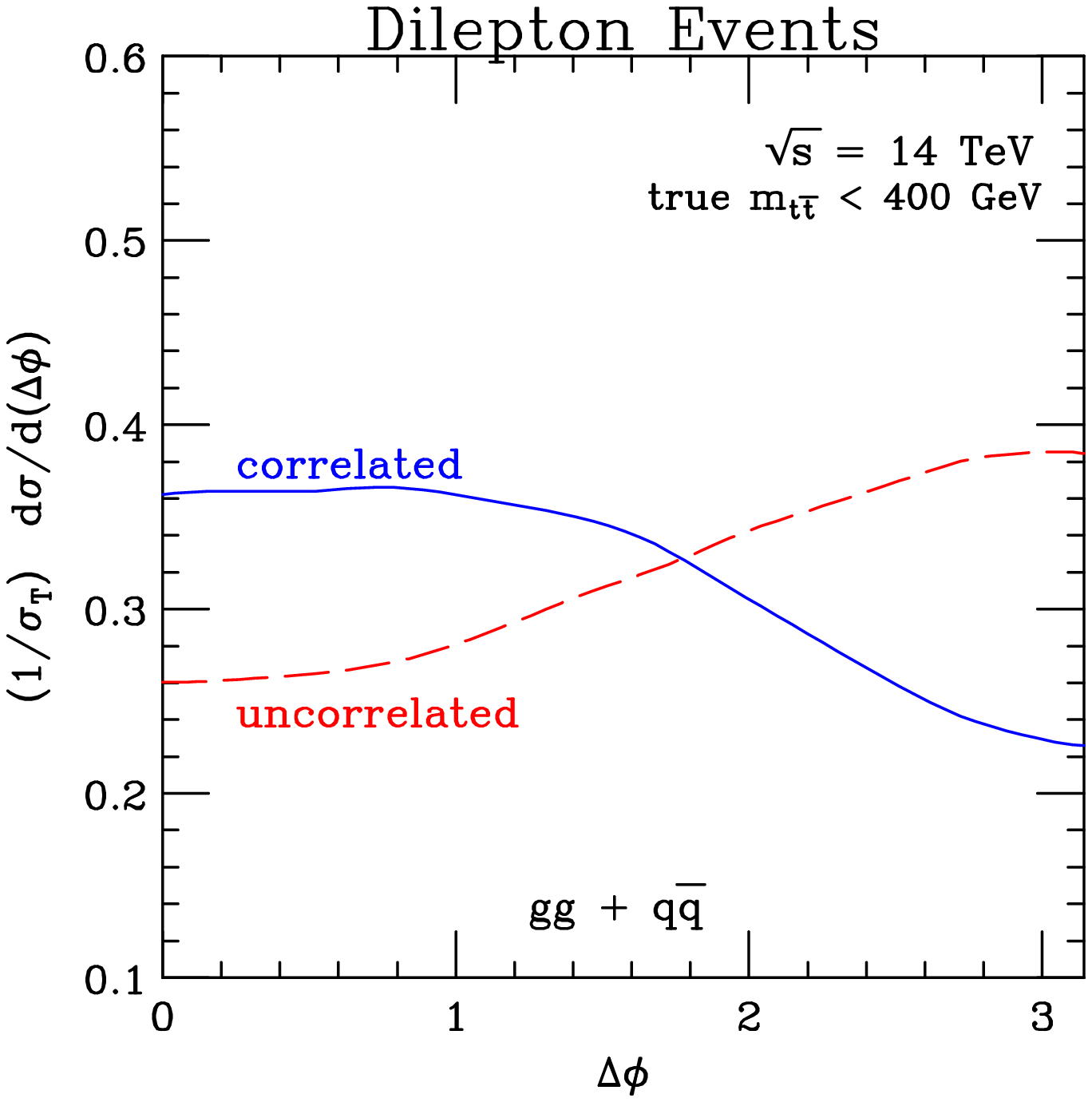}
\vspace*{9.5cm}
\caption{The differential distribution of $\Delta\phi$, 
$(1/\sigma_T)~ d\sigma/d(\Delta\phi) $.
The solid curve is for the fully correlated case
whereas the dashed curve assumes that the top quarks decay
spherically in their respective rest frames.
A cut restricting the invariant mass of the $t\bar{t}$ pairs
to a maximum of 400~GeV has been applied to these distributions.
}
\label{fig:azim-truecut}
\end{figure*}

In Fig. \ref{fig:azim-truecut} we have plotted the  
$\Delta \phi$ distribution in
the dilepton channel for $t\bar{t}$ production incorporating a cut 
which restricts
the true invariant mass of the $t\bar{t}$ pair
to less than 400 GeV.  This plot shows
results for both
the fully-correlated and the uncorrelated matrix elements 
including both $gg \rightarrow t\bar{t}$ and
$q\bar{q} \rightarrow  t\bar{t}$ channels.  
A clear distinction between 
the correlated and uncorrelated 
decays\footnote{The corresponding $\Delta \eta$ distribution shows 
almost no difference between
correlated and uncorrelated matrix elements. Thus all the difference 
in the $\Delta R$ distributions comes from the $\Delta\phi$ 
distributions.}
is seen in this figure:  the difference 
between the two $\Delta \phi$ distributions is about 40\% at both 
$\Delta \phi=0$ (enhancement) and $\Delta \phi =\pi$ (suppression).
With this cut, 10\% of the total cross section for $t\bar{t}$ 
production survives at leading order.
   
Unfortunately, the presence of the two neutrinos in the final state
of dilepton events complicates the selection of events.
The available kinematic constraints leave an up to 8-fold 
ambiguity\footnote{The presence of a pair of quadratic constraints
in the kinematic equations leads to up to 4 solutions for any
given pairing of the $b$ jets with the two charged leptons.  
Since
there are two possible pairings, as many as 8 different
solutions could result.
However, not all of these solutions need be real, and so there
are often fewer than the maximum possible number of solutions.}
in the reconstruction of the neutrino momenta from the 
available observed momenta and energies.
Given the ease of measuring the azimuthal angles of charged leptons, 
it is worthwhile to investigate an alternative to the true
$t\bar{t}$ invariant mass cut.  


\begin{figure*}[tbh]
\includegraphics{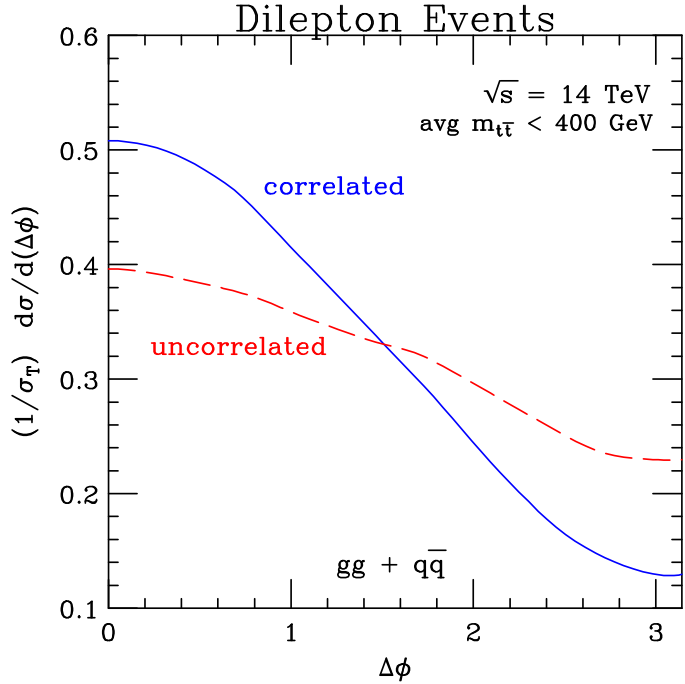}
\vspace*{9.5cm}
\caption{The differential distribution of $\Delta\phi$, 
$(1/\sigma_T)~ d\sigma/d(\Delta\phi) $.
The solid curve is for the fully correlated case
whereas the dashed curve assumes that the top quarks decay
spherically in their respective rest frames.
A cut restricting the average reconstructed
invariant mass of the $t\bar{t}$ pairs
to a maximum of 400~GeV has been applied to these distributions.
}
\label{fig:azim-avgcut}
\end{figure*}

The simplest option one could imagine is to simply take the 
(na\"\i ve) unweighted average 
$\langle m_{t\bar{t}} \rangle$
of all of the real solutions returned
by the neutrino reconstruction algorithm.
In Fig.~\ref{fig:azim-avgcut}
we present the results of implementing just that option:
the cut used to generate this figure 
requires that $\langle m_{t\bar{t}}\rangle$ be less than 400~GeV.
With this cut approximately 5\% of the total cross section 
for $t\bar{t}$ production survives at leading order.
This is smaller than the
fraction passing a 400~GeV cut on the true value of $m_{t\bar{t}}$
since only those events where all the spurious solutions are 
sufficiently small will survive.  On the other hand, the sample
passing this cut will contain a few events where the true value
of $m_{t\bar{t}}$ is above 400~GeV, but, because the spurious
solutions produced smaller values, the average was below 400 GeV.
Turing to the $\Delta\phi$ distribution and comparing 
to the cut on the true value of $m_{t\bar{t}}$, one
sees a rather large effect on the shape of the distributions.
However, this effect (an enhancement near $\Delta\phi=0$ and a
depletion near $\Delta\phi=\pi$) occurs for {\it both}\ the
correlated and uncorrelated data sets.  Thus the difference
between the two distributions remains at roughly the 40\% level.
No effort has been to optimize this invariant mass cut.  
Perhaps there are other variables that will
do better than unweighted average 
$\langle m_{t\bar{t}} \rangle$, or perhaps
400 GeV is not the optimal cut value.  
Nevertheless,
what we have here is a proof-in-principle
that these correlations can be measured in an experiment.


\subsection{ZMF $\cos\theta$ Distribution for Lepton-plus-jets Events}

Turning to the lepton-plus-jets channel, we have found that the
cosine of the opening angle between the charged lepton and the
$d$-quark jet as viewed in the zero momentum frame (ZMF) is 
sensitive to the presence or absence of correlations between production
and decay [see also the discussion near Eq.~(\ref{eqn:ratio})].  
For this type of event the kinematic constraints
provide more equations than unknowns.
Thus, the ZMF may be reconstructed
without ambiguity more than 98\% of the time by discarding 
those solutions which do not pass some rudimentary quality-control
cuts:  the neutrino energy ought to be positive
in a correctly reconstructed event; futhermore, the neutrino
and top quark mass-shell constraints ought to be satisfied to
sufficient accuracy.\footnote{Because of jet energy measurement 
uncertainties,
it is not expected that a correctly reconstructed top quark or
neutrino will be precisely on mass shell.  The 
exact definition of
``sufficient precision'' is therefore a detector-dependent issue to be 
determined by the experimental collaborations.
} 
Discarding the ${\sim}2\%$ of events that have 
more than one viable reconstruction of the ZMF is an acceptible option.

\begin{figure*}[tbh]
\includegraphics{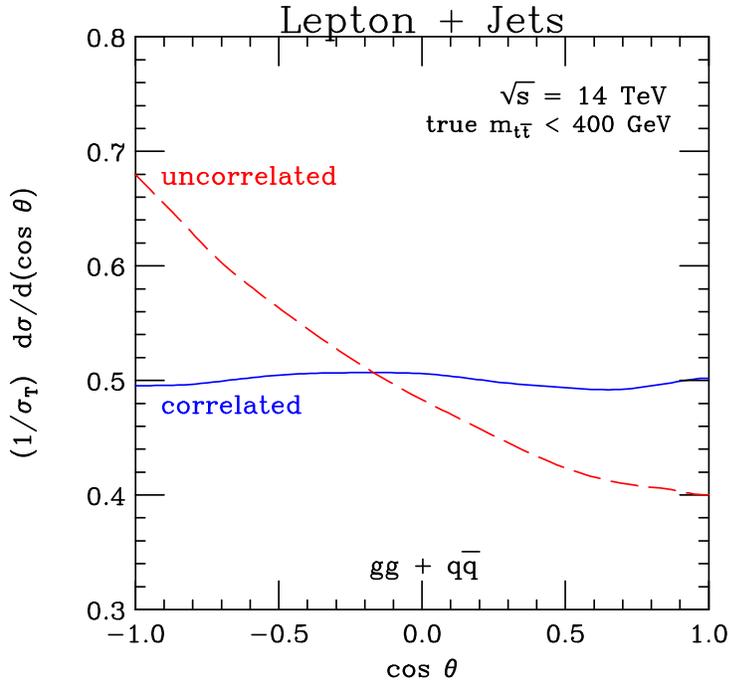}
\vspace*{9.5cm}
\caption{The differential distribution of $\cos\theta$, 
$(1/\sigma_T)~ d\sigma/d(\cos\theta) $, where $\theta$ is the
ZMF angle between the charged lepton and the $d$-quark jet
(defined to be the jet which is spatially the closest to the
$b$-tagged jet in the $W$ rest frame; this is also the jet
with the lowest energy in the top quark rest frame).
The solid curve is for the fully correlated case
whereas the dashed curve assumes that the top quarks decay
spherically in their respective rest frames.
A cut restricting the 
invariant mass of the $t\bar{t}$ pairs
to a maximum of 400~GeV has been applied to these distributions.
}
\label{fig:ZMFcth}
\end{figure*}

For the purposes of generating this distribution, we define the
$d$-quark jet to be the jet which is spatially the closest to the
$b$-tagged jet in the $W$ rest frame, as was used 
in Ref.~\cite{mahlon and parke}. 
This is equivalent to using the lowest energy
jet in the top quark rest frame as advocated in 
Ref.~\cite{Shelton:2008nq}.
Fig.~\ref{fig:ZMFcth} displays the results for this distribution
using only those events that pass the cut $m_{t\bar{t}} < 400$~GeV.
For fully-correlated top decays this distribution is nearly flat,
whereas for spherical decays there is a strong peaking
near $\cos\theta = -1$.  That is, because of the correlations between
production and decay, the lepton and the $d$-jet
tend to be significantly less back-to-back in the ZMF than if
no such correlations were present.
Approximately 9\% of the total cross section for $t\bar{t}$ 
production survives this cut at leading order,
even at reduced center-of-mass energy.


\subsection{Varying the Energy of the LHC}


\begin{figure*}[tbh]
\includegraphics{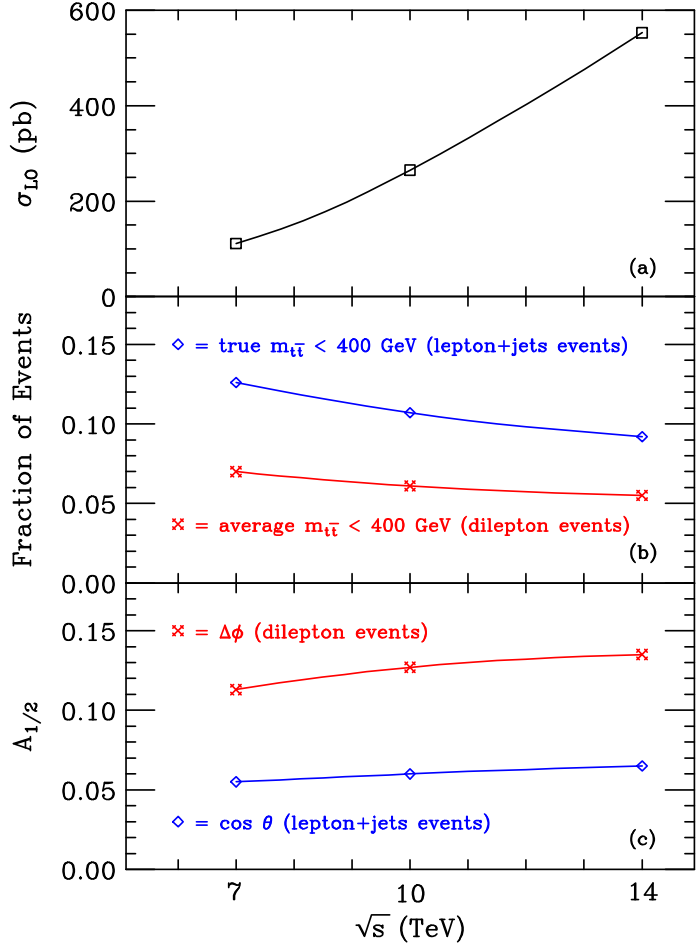}
\vspace*{12.5cm}
\caption{Effects of varying the machine center-of-mass
energy $\sqrt{s}$.
(a) Total leading order cross section for $pp\rightarrow t\bar{t}$.
These values should be multiplied by the branching fraction to
dileptons (4.6\%) or lepton plus 
jets (29\%), as appropriate.  We include only the
$e$ and $\mu$ channels.
(b)~Fraction of 
dilepton and lepton plus events with $m_{t\bar{t}} < 400$~GeV.  
For dilepton
events (crosses) we employ the unweighted average of the up to 8
solutions for the $t\bar{t}$ invariant mass.  For lepton+jets events
(diamonds)
the true value of $m_{t\bar{t}}$ may be reconstructed and used in 
event selection.
(c)~Half of the area between the appropriate unit-normalized angular
distributions for the fully correlated and spherical cases.
For lepton+jets events (crosses), we use the distribtuion in 
$\cos\theta$,
where $\theta$ is the angle between the charged lepton and the
$d$-jet candidate in the zero momentum frame of the event.
For dilepton events (diamonds), we use
the azimuthal opening angle $\Delta\phi$ between 
the two charged leptons.
}
\label{fig:hybrid}
\end{figure*}
So far we have used 14 TeV for the energy of the LHC. 
However, it is now clear that the LHC will not reach this energy 
for a number of years so we have investigated what happens 
for a reduced center-of-mass energy in this section.
The results we describe below are summarized in Fig.~\ref{fig:hybrid}.

The primary result of a lower center-of-mass energy is a big reduction
in the $t\bar{t}$ production cross section because of the
reduced gluon luminosity.  Panel~(a) in Fig.~\ref{fig:hybrid}
tracks the leading order cross section 
from 7 to 14~TeV center-of-mass energy.  We see that a factor
of 2 reduction in $\sqrt{s}$ produces a reduction of about a factor
of 5 in the $t\bar{t}$ production cross section.  Panel~(b) 
illustrates the fact that the fraction of 
dilepton and lepton plus jet events surviving 
the $m_{t\bar{t}}$ or 
$\langle m_{t\bar{t}}\rangle$
cut we advocate does not change very much as $\sqrt{s}$ is varied 
between 7~and 14~TeV.  
Finally, in panel~(c) of each figure we compare the quantity
${\cal A}_{1/2}$, which is defined as half of the area between the
correlated and uncorrelated predictions for the unit-normalized
angular distributions.  This quantity ranges from 0 in the 
case where the distribution is completely independent of whether or
not correlations are present up to 1 in the case where the two 
distributions (correlated decays or spherical decays) do not overlap
at all.  In both channels, there is relatively little dependence
on $\sqrt{s}$ for this measure of the difference between the correlated
and spherical cases.
Thus, the biggest issue related to the observation of these spin 
correlations at the LHC running at reduced energy comes from the
greatly diminished cross section:  the correlations themselves
remain at roughly the same level.  Fortunately, even a reduction of
the number number of $t\bar{t}$ pairs estimated in the introduction
by a factor of 5 leaves ample statistics to hope for at least
a preliminary observation of these spin correlation effects,
even at reduced center-of-mass energy.


\subsection{NLO Effects}

Higher-order QCD effects enhance the total cross section
especially near threshold.  However, previous studies 
on $e^+ e^- \rightarrow t\bar{t}$ demonstrate that such 
corrections to the spin correlations are small, 
see Ref.~\cite{Kodaira:1998gt}.
One can understand 
this physically since the emission of soft gluons from a top quark 
cannot flip the spin of the top quark.  We have done a preliminary
study of the NLO effects using both MCFM, \cite{MCFM} and MC@NLO, \cite{Frixione:2007zp}, incorporating a cut 
which restricts
the invariant mass of the $t\bar{t}$ pair
to be less than 400 GeV. Both these Monte Carlos show that the
tree-level effects discussed earlier in this paper are also present 
at NLO.\footnote{In  Fig. 1 of Ref.~\cite{Frixione:2007zp} one can 
see the size of the correlations without the invariant mass cut.}  
We need not check the Monte Carlo of Ref.~\cite{Melnikov:2009dn}
since it agrees with the other two Monte Carlos for the correlated decays.
A detailed study at NLO where the invariant mass of the $t\bar{t}$ 
pair is reconstructed from the decay products is beyond the scope of 
this work. A NLO Monte Carlo with a switch that allows the user to 
toggle between fully correlated top quark decays and spherical decays 
of the tops would be very useful for such a study.  However, no such 
Monte Carlo exists at present.


\section{Summary and Conclusions}\label{sec:conclude}
In this paper we have shown how to observe spin correlations in top 
quark pair production at the LHC. To our surprise, the observation of 
these correlations is easier at the LHC than at the Tevatron.  The 
reason for this is that at the LHC the dominant production mechanism 
for top quark pairs is gluon-gluon fusion, which at low $\hat{s}$ is 
dominated by the fusion of like helicity pair gluons.  The fusion of 
like helicity gluons produces top quark pairs in a LL or RR helicity 
configuration. When such top quarks decay, they produce charged 
leptons which possess very strong 
azimuthal correlations.  These correlations 
can be easily seen in the laboratory frame once a cut on the 
invariant mass of the top quark pair is made.  There is not need to 
reconstruct the top quark rest frame as is required to see the 
correlations of top quark pairs at the Tevatron.  The analysis has 
been extended to the lepton plus jets channel by ``identifying''
which jet is the $d$-quark jet from the $W$-boson decay 
on a statistical basis.  Apart from the reduction of the total cross 
section, the size of these spin correlations is approximately 
independent of the energy of the LHC in the 7 to 14 TeV range.
Thus, we expect that these effects could be observed in early 
running of the LHC.

\acknowledgments

SP would like to thank Jiro Kodaira for many enlightened
discussions over the course of many years.  John Campbell,
Keith Ellis, Stefano Frixione and Brian Webber
are acknowledged for fruitful discussions about the NLO comparison.  
SP
also thanks the CERN Theory group for hospitality during the "Top Quark
Physics Institute" during May of 2009.  GM would like to thank
the Fermilab Theory group for its kind hospitality during multiple
summer visits to concentrate on this work.
Fermilab is operated by the Fermi Research Alliance under contract 
no.~DE-AC02-07CH11359 with the U.S. Department of Energy.



\appendix

\section{Universal Properties of the Spin Amplitudes}
In this Appendix we demonstrate the properties of the spin amplitudes 
and their relationship to the Wigner $d$-functions and the helicity 
amplitudes.  Of particular importance is the fact that spin 
amplitudes at either end of the spin chain contain all the 
information necessary to obtain all the other spin amplitudes by 
simple algebraic manipulations. This fact was used extensively 
earlier in this paper.

\subsection{Wigner $d$-functions and connection 
to the 
spin amplitudes}

The amplitude for production of a single massive particle
with spin $j$ and spin-projection 
$m$ in the generalized spin basis 
described in Sect.~\ref{sec:gen-basis} and illustrated in 
Fig.~\ref{fig:s-vector} may be written
as a linear superposition of the matrix elements of the rotation
operator~\cite{Jacob+Wick}:
\beqa
{\cal A}_{j,m}(\xi) &\equiv& \langle 
\overrightarrow{x} \vert j \ns m\rangle_{\xi}
\cr &=&
\sum_{m'} d^j_{m,m'}(-\xi) H_{m'}.
\label{d-exp}
\eeqa
The $d^j_{m,m'}$ appearing in Eq.~(\ref{d-exp}) are the
Wigner $d$-functions, chosen to conform to the conventions
of Rose~\cite{Rose}.  In particular, we write $-\xi$ since our
angle $\xi$ is measured in the clockwise direction whereas 
the angle $\beta$ in Ref.~\cite{Rose} is counterclockwise.
Since 
\beq
d_{m,m'}^j(0) = \delta_{m,m'},
\eeq
we see that
\beq
H_m = {\cal A}_{j,m}(0).
\eeq
That is, the coefficients $H_m$ are 
the conventional helicity
amplitudes for the process with a particular choice of 
relative phases.

Similarly, the production of a pair of massive particles (spins
$j_1$ and $j_2$, spin projections $m_1$ and $m_2$ along 
independent spin
axes oriented at the clockwise angles $\xi$ and $\xi'$ 
with respect to the recoil direction in the scattering plane) may 
be decomposed as
\beqa
{\cal A}_{j_{{}_1},m_{{}_1},j_{{}_2},m_{{}_2}}(\xi,\xi') 
= \sum_{m} \sum_{m'} 
d^{j_{{}_1}}_{m_{{}_1},m}(-\xi) 
d^{j_{{}_2}}_{m_{{}_2},m'}(-\xi') 
H_{m,m'}.
\label{double-d-exp}
\eeqa
The extension to more than two particles in the final state
or to spin axes which point out of the production plane, involving
the introduction of the Wigner $D$-functions in place
of the (simpler) $d$-functions, is straightforward, but beyond
the scope of this appendix.

\subsection{$\xi\rightarrow\xi\pm\pi$ rule}

Intuitively,  we expect that the probability for producing
spin projection $+m$ along some axis ought to be equal
to that for producing spin projection $-m$ along  
minus that axis:
\beq
\vert {\cal A}_{j,m}(\xi\pm\pi) \vert= 
 \vert {\cal A}_{j,-m}(\xi)\vert.
\label{xi2xi+pi}
\eeq
That is,
flipping the $z$ component of the spin and rotation by $\pm \pi$
give the same results up to a phase.  This feature of the spin
amplitudes is a consequence of the properties
of the Wigner $d$-functions appearing
in Eqs.~(\ref{d-exp}) and~(\ref{double-d-exp}).


\subsection{Differential relations among the spin amplitudes}
Consider a state with total spin $j$ and projection $j_z =m$
in the general spin basis illustrated in Fig.~1.
Applying the rotation operator 
converts this to a state where the spin 
axis is at $\xi + \Delta\xi$ instead~\cite{Jacob+Wick}:
\beq
{\cal A}_{j,m}(\xi+\Delta\xi)
= e^{i\Delta\xi J_y} {\cal A}_{j,m}(\xi).
\label{J144}
\eeq
In the limit $\Delta\xi \rightarrow 0$ Eq.~(\ref{J144})
may be rewritten as
\beq
{ {\dee} \over {\dee\xi} }
{\cal A}_{j,m}(\xi)
= {1\over2} (J_{+} - J_{-}) {\cal A}_{j,m}(\xi).
\label{master}
\eeq
In this expression, we have replaced $J_y$ by the appropriate
linear combination of raising and lowering operators.
Thus
\beqa
{ {\dee} \over {\dee\xi} }
{\cal A}_{j,m}(\xi)
&=& {1\over2} \sqrt{j(j+1)-m(m+1)} \ts {\cal A}_{j,m+1}(\xi)
\cr
&-& {1\over2} \sqrt{j(j+1)-m(m-1)} \ts {\cal A}_{j,m-1}(\xi)
\label{Master}
\eeqa

\subsection{Starting at top or bottom of spin chain}
It is useful to record the explicit results
of applying Eq.~(\ref{Master})
to the ends of the spin chain:
\beq
{\cal A}_{j,\pm(j-1)}(\xi) = \mp\sqrt{2\over{j}} \ts 
{{\partial}\over{\partial\xi}}
{\cal A}_{j,\pm j}(\xi).
\label{OneLink}
\eeq
A second differentiation allows us to conclude that
\beq
{\cal A}_{j,\pm(j-2)}(\xi) =
{{{1}\over{\sqrt{j(2j{-}1)}}}
\Biggl[ j + 2 {{\partial^2}\over{\partial\xi^2}}} \Biggr]
\ts{\cal A}_{j,\pm j}(\xi).
\label{TwoLink}
\eeq
This process could be repeated as many times as required.  However,
when used in conjunction with the $\xi\rightarrow \xi\pm\pi$ 
rule of~Eq.~(\ref{xi2xi+pi}), Eqs.~(\ref{OneLink}) and~(\ref{TwoLink})
allow the calculation of all of the amplitudes in the spin chain
for spins up to and including $5\over2$, starting from either end
of the chain ($m=\pm j$).  The relative simplicity of the operations
involved in these relations makes for a substantial computational
savings over the calculation of the complete set of amplitudes
by direct means.   Of course, once you have all the spin 
amplitudes then
the helicity amplitudes can be easily obtained, including 
relative phases, by setting $\xi =0$.

Thus, the spin amplitudes at the top and bottom of the spin chain 
contain all the information about a given process and all 
other spin amplitudes and helicity 
amplitudes can be derived from them.
This special property of these spin amplitudes is simply reflected 
in the Wigner $d$-functions, 
$d^j_{j,m}$ for  $m=\{-j,\cdots,j\}$, which 
form a set of (2j+1) linearly independent functions~\cite{edmonds}.

\subsection{Examples:}
An explicit illustration of the relationships contained in 
Eqs.~(\ref{OneLink}) and~(\ref{TwoLink}) is provided by the
$gg\rightarrow t\bar{t}$ process considered in this paper.
After setting $\xi'=\xi$ (back-to-back spin axes in the ZMF), we
can organize the amplitudes according to the total spin in
the final state.  In doing this we need to keep in mind that
for this choice of spin axes, {\it unlike}\ spin $t\bar{t}$ pairs
have spins that point in the same spatial direction. 
Thus 
\beq
\vert {\cal A}_{1,1}(\xi) \vert =
\vert {\cal A}(gg\rightarrow t_\up \bar{t}_\down) \vert
\eeq
and
\beq
\vert {\cal A}_{1,-1}(\xi) \vert =
\vert {\cal A}(gg\rightarrow t_\down \bar{t}_\up) \vert.
\eeq
The $\xi$-dependent linear combination of the $\up\up$ and $\down\down$
amplitudes must be ${\cal A}_{1,0}$; the orthogonal ($\xi$-independent)
combination is ${\cal A}_{0,0}$:
\beq
\vert {\cal A}_{1,0}(\xi) \vert =
{{1}\over\sqrt{2}}
\vert
 {\cal A}(gg\rightarrow t_\up \bar{t}_\up) 
 + {\cal A}(gg\rightarrow t_\down \bar{t}_\down) \vert,
\eeq
\beq
\vert {\cal A}_{0,0}(\xi) \vert =
{{1}\over\sqrt{2}}
\vert
 {\cal A}(gg\rightarrow t_\up \bar{t}_\up) 
 - {\cal A}(gg\rightarrow t_\down \bar{t}_\down) \vert.
\eeq
For like-helicity gluons Eqs.~(31) and (32) lead to
\beqa
\vert{\cal A}_{1,1}(\xi)\vert &\sim&
\gamma^{-1} \vert \beta\sxi \vert
\cr
\vert{\cal A}_{1,0}(\xi)\vert &\sim& 
\sqrt{2}\gamma^{-1} \vert \beta\cxi \vert
\qquad \hbox{and} \qquad
\vert{\cal A}_{0,0}(\xi)\vert \sim \sqrt{2}\gamma^{-1}.
\cr
\vert{\cal A}_{1,-1}(\xi)\vert &\sim&
\gamma^{-1} \vert \beta\sxi \vert
\label{likeglue}
\eeqa
The three $j=1$ amplitudes in~(\ref{likeglue})
satisfy
\beq
\vert {\cal A}_{1,0}(\xi) \vert =
\Bigl\vert \sqrt{2} {\partial\over{\partial\xi}} {\cal A}_{1,1}(\xi)
\Bigr\vert
\label{OneZero}
\eeq
and
\beq
\vert {\cal A}_{1,-1}(\xi) \vert =
\Biggl\vert 
\Bigl\{ 1 + 2 { {\partial^2}\over{\partial\xi^2} } \Bigr\}
{\cal A}_{1,1}(\xi)
\Biggr\vert
\label{OneOne}
\eeq
as implied by Eqs.~(\ref{OneLink}) and~(\ref{TwoLink}).

A different realization of these relationships is provided by
the unlike-helicity gluons for which Eqs.~(22) and~(23) lead to
\beqa
\vert{\cal A}_{1,1}(\xi)\vert &\sim&
\beta\sth ( 1+\cth\cxi+\gamma^{-1}\sth\sxi )
\cr
\vert{\cal A}_{1,0}(\xi)\vert &\sim& 
\sqrt{2}\beta\sth\vert \cth\sxi-\gamma^{-1}\sth\cxi \vert
\qquad\hbox{and}\qquad
\vert{\cal A}_{0,0}(\xi)\vert = 0.
\cr
\vert{\cal A}_{1,-1}(\xi)\vert &\sim&
\beta\sth ( 1-\cth\cxi-\gamma^{-1}\sth\sxi )
\label{oppoglue}
\eeqa
This is identical (up to overall factors) to what happens for
the processes
$q\bar{q}\rightarrow t\bar{t}$
and similar to what happens for
$e^{+}e^{-}\rightarrow t\bar{t}$~\cite{Parke:1996pr}.
The spin amplitudes for 
$e^{+}e^{-}\rightarrow Zh$~\cite{ref:deconstruction,ZHZA}
also satisfy Eqs.~(\ref{OneZero}) and~(\ref{OneOne}).

Finally, we have verified that 
the processes
$e^{+}e^{-}\rightarrow W^{+}W^{-}$
and
$ZZ$
provide examples of the $j=2$ versions of Eqs.~(\ref{OneLink})
and~(\ref{TwoLink}).
Indeed, the derivative relations between the spin amplitudes 
for these processes noted
in Ref.~\cite{ref:deconstruction} are a direct consequence of
Eq.~(\ref{OneLink}).


\section{The process $q\bar{q}\rightarrow t\bar{t}$}

For completeness we give here the matrix element squared for
$q\bar{q}\rightarrow t\bar{t}$ with the subsequent decay of the top
quarks. Starting from Eqn.~(\ref{eqn:qqbar-spinor}) and using the
substitutions given in Eq.~(\ref{eqn:add-decays}),
it is easy to add the decays of the on-mass-shell top quarks
($t\rightarrow b + \bar{e} + \nu$ and
$\bar{t}\rightarrow \bar{b} + \mu + \bar\nu$).
Thus, the total matrix element squared for top quark production
and decay via quark-antiquark annihilation, summed over the colors of
the
incoming and outgoing quarks, is given by
\begin{eqnarray}
\vert {\cal A}\vert ^2_{RL}  + \vert {\cal A}\vert ^2_{LR}  
&=&{\cal K}_{q\bar{q}}
\Biggl\{ 2(t \cdot \bar{e})(\bar{t} \cdot \mu)
\Bigr[(q\cdot t)^2+(\bar{q} \cdot t)^2 \Bigl] \nonumber 
\cr &&
-m_t^2\Bigl[(q\cdot \bar{q}) \Bigl( (t \cdot \bar{e})(t\cdot \mu)
+(\bar{t}\cdot \bar{e})(\bar{t} \cdot \mu)-m_t^2(\bar{e} \cdot \mu)
\Bigr)
\nonumber 
\cr && \qquad
-2 \Bigl((q \cdot t)(q \cdot \mu) (\bar{q} \cdot \bar{e}) 
+(\bar{q} \cdot t)(q \cdot \bar{e})(\bar{q} \cdot \mu) 
-(q \cdot t)(\bar{q} \cdot t)(\bar{e} \cdot \mu)\Bigr)\Bigr]\Biggr\}. 
\nonumber \\
\label{eqn:qqbar-full}
\end{eqnarray}
This has the same functional form as the part of 
Eqn.~(\ref{eqn:unlike-full}) in the second set of curly brackets.
The overall factor ${\cal K}_{q\bar{q}}$ is given by
\begin{eqnarray}
{\cal K}_{q\bar{q}} = {{2^6 g_s^4}\over{(q\cdot\bar{q})^2}}
{{g_w^8}\over{(m_t\Gamma_t)^4}}
{{b\cdot\nu}\over{(2\bar{e}\cdot \nu-m^2_w)^2 + (m_w \Gamma_w)^2}}
{{\bar{b}\cdot\bar\nu}\over{(2\mu \cdot \bar{\nu}-m^2_w)^2 + (m_w \Gamma_w)^2}}. 
\label{eqn:qqKfactor}
\end{eqnarray}
Eqs.~(\ref{eqn:qqbar-full})~and (\ref{eqn:qqKfactor}) are the 
Lorentz-invariant equivalents of 
Eqs.~(4) and (5) of Ref.~\cite{Mahlon:1997uc}.


\end{document}